\UseRawInputEncoding
\documentclass[journal=jpcbfk,manuscript=article]{achemso}
\usepackage{longtable}
\usepackage{multirow}
\usepackage{amsmath}
\usepackage{subfigure}
\usepackage[usenames,dvipsnames]{color}
\usepackage{soul}
\usepackage{threeparttable}
\usepackage{xr}
\usepackage{graphicx}
\usepackage{amsmath,amssymb}
\usepackage{caption}
\usepackage{color}
\usepackage{dcolumn}
\usepackage{bm}
\usepackage{float}
\usepackage{subfig}
\usepackage{textcomp}
\usepackage{url}
\usepackage[normalem]{ulem}

\makeatletter
\newcommand*{\addFileDependency}[1]{
  \typeout{(#1)}
  \@addtofilelist{#1}
  \IfFileExists{#1}{}{\typeout{No file #1.}}
}
\makeatother
 
\newcommand*{\myexternaldocument}[1]{%
    \externaldocument{#1}%
    \addFileDependency{#1.tex}%
    \addFileDependency{#1.aux}%
}
\myexternaldocument{si}

\author{Seyedeh Maryam Salehi} \affiliation[University of
  Basel]{Department of Chemistry, University of Basel,
  Klingelbergstrasse 80 , CH-4056 Basel, Switzerland.}  \author{Markus
  Meuwly} \affiliation[University of Basel]{Department of Chemistry,
  University of Basel, Klingelbergstrasse 80 , CH-4056 Basel,
  Switzerland.}  \email{m.meuwly@unibas.ch}

\title {Site-Selective Dynamics of Ligand-Free and Ligand-Bound
  Azidolysozyme}

\begin{document}

\begin{abstract}
Azido-modified alanine residues (AlaN$_3$) are environment-sensitive,
minimally invasive infrared probes for the site-specific investigation
of protein structure and dynamics. Here, the capability of the label
is investigated to query whether or not a ligand is bound to the
active site of Lysozyme and how the spectroscopy and dynamics change
upon ligand binding. The results demonstrate specific differences for
center frequencies of the asymmetric azide stretch vibration, the long
time decay and the static offset of the frequency fluctuation
correlation function - all of which are experimental observables -
between the ligand-free and the ligand-bound, N$_3$-labelled
protein. Changes in dynamics can also be mapped onto changes in the
local and through-space coupling between residues by virtue of
dynamical cross correlation maps. This makes the azide label a
versatile and structurally sensitive probe to report on the dynamics
of proteins in a variety of environments and for a range of different
applications.
\end{abstract}

\noindent
Proteins are essential for function and sustaining life of
organisms. Experimentation and computational studies have clarified
that protein function involves both, structure {\it and}
dynamics.\cite{schuler:2017,zhou:2016,zhang:2019} However,
characterizing structural and functional dynamics of proteins at the
same time under physiological conditions in the condensed phase, which
is prerequisite for understanding cellular processes at a molecular
level, remains a challenging undertaking.\cite{schuler:2017}
Vibrational spectroscopy, in particular 2-dimensional infrared (2D IR)
spectroscopy, has been shown to be a powerful tool for studying the
structural dynamics of various biological
systems\cite{2DIRbook-Hamm-2011}. One of the particular challenges is
to obtain structural and environmental information in a site-specific
manner. To address this, significant effort has been focused on the
development and application of various infrared (IR)
reporters\cite{gai:2015,cho:2013} that absorb in the frequency range
of 1700-2800 cm$^{-1}$ to discriminate the signal from the strong
protein background.\cite{gai:2011,hamm.rev:2015} Such IR probes have
provided valuable information about the structure and dynamics of
complex systems. For example, nitrile probes have helped to clarify
the role of electrostatic fields in enzymatic
reactions\cite{boxer:2014,hammes-schiffer:2013} or to elucidate the
mode of drug binding to proteins.\cite{hochstrasser:2013,blasie:2009}
Isotope edited carbonyl spectroscopy was used to characterize the
mechanism of protein folding and amyloid
formation\cite{dayer:2007,zanni:2012} or the structure and function of
membrane
proteins\cite{hochstrasser:2011,hochstrasser.science:2011}. Additional
molecular groups such as thiocyanate,\cite{bredenbeck:2014}
cyanamide,\cite{cho:2018} sulfhydryl vibrations of
cysteines,\cite{hamm:2008} deuterated carbons,\cite{romesberg:2011}
carbonyl vibrations of metal-carbonyls, \cite{zanni:2013}
cyanophenylalanine,\cite{thielges:2019} and azidohomoalanine
(Aha)\cite{hamm:2012} have also been explored.\\

\noindent
In the present work AlaN$_3$, an analogue of azidohomoalanine (Aha)
that has been shown to sensitively report on local structural changes
while still being minimally invasive,\cite{hamm.rev:2017,MS.lys:2021}
is used as the probe. This modification can be incorporated into
proteins at virtually any position via known expression
techniques.\cite{bertozzi:2002} The asymmetric stretch frequency of
-N$_3$ is at $\sim 2100$ cm$^{-1}$ and has a reasonably high
extinction coefficient of 300-400 M$^{-1}$cm$^{-1}$ which makes it an
ideal spectroscopic reporter.\cite{hamm:2012} Aha has been used for
biomolecular recognition after incorporation into the peptide
directly\cite{hamm:2012,hamm.rev:2017} or in the vicinity of the
binding area of a PDZ2 domain\cite{stock:2018}, to detect the
water-specific response of azide vibrations when attached to small
organic molecules\cite{londergan:2012}, or to probe the frequency
shift and fluctuation due to its sensitivity to the local
electrostatic environments and dynamics.\cite{cho:2008,MS.lys:2021}
Such studies confirm that AlaN$_3$ and/or Aha are
environment-sensitive IR probes and suitable modifications for
site-specific investigations of protein structure.\\

\noindent
With its picosecond time resolution, IR spectroscopy provides direct
information about the structural dynamics around a probe molecule with
high temporal resolution.\cite{2DIRbook-Hamm-2011,thielges:2016}
Moreover, introducing IR probes with isolated vibrational frequencies
overcomes the problem of spectral congestion that complicates
discrimination and analysis of desired vibrational bands. With that,
the inter- and intramolecular coupling between degrees of freedom or
the local structure or dynamics of biological systems can be
specifically probed and characterized. Such an approach relies on the
sensitivity of the probe to report on changes in the vibrational
frequencies induced by alterations in the local electrostatic
interactions in the vicinity of the probe.\cite{thielges:2019}\\

\noindent
IR spectroscopy is a potentially advantageous technique to
characterize ligand binding to
proteins.\cite{Suydam.halr.sci.2006,MM.lys:2017} Its success depends
in part on the notion that when a ligand binds to a protein, the
frequency of an infrared active vibration shifts due to the different
electric field in solution - often water - and in the protein binding
site. Such an approach often requires the ligand to be modified,
e.g. through addition of a suitable label such as -CN as in
benzonitrile. This has been successfully demonstrated for benzonitrile
in the active site of WT and mutant lysozyme.\cite{MM.lys:2017}\\

\noindent
Alternatively, the protein can be selectively modified by attaching a
spectroscopic label at strategic positions so that the binding process
and functional dynamics can be interrogated with the functional and
unmodified ligand. This has the potential advantage that interactions
between the ligand and the surrounding protein are unaltered. These
interactions contribute the majority of the enthalpic part to the
binding free energy and therefore directly affect the affinity of the
ligand and its rate of unbinding. In the present work changes in 1d-
and 2d-IR signatures of the azido group attached to all alanine (Ala)
residues of Lysozyme upon binding of cyano-benzene (PhCN) are
determined. In addition, the changes of the environmental dynamics
around all AlaN$_3$ are quantified for ligand-free vs. ligand-bound
Lysozyme. Such differences are experimentally observable and yield
valuable insight into the energetics and dynamics of protein-ligand
binding.\\

\begin{figure}[H]
\begin{center}
\includegraphics[width=0.8\textwidth]{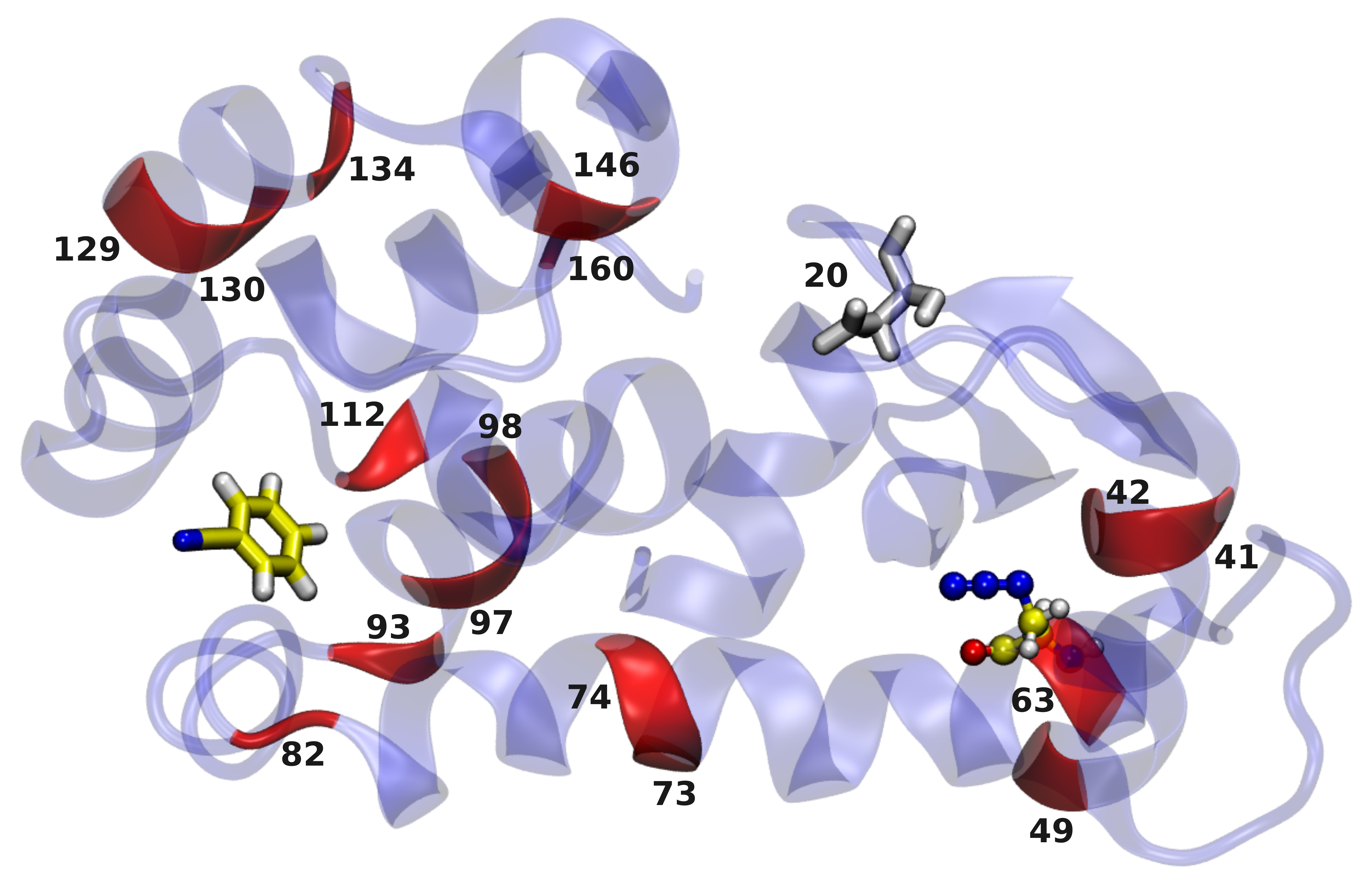}
\caption{The structure of Lysozyme including PhCN (Licorice) in the
  active site and Ala63N$_3$ (CPK) as an example of attaching the
  -N$_3$ label. The Ala residues at positions 41, 42, 49, 63, 73, 74,
  82, 93, 97, 98 , 112, 129, 130, 134, 146, 160 (red NewCartoon) are
  labelled one at a time. The rest of the protein is shown as blue
  NewCartoon except for residue Asp20 which is in white Licorice.}
\label{fig:lys}
\end{center}
\end{figure}

\noindent
The dynamics of WT Lysozyme without and with labeled alanine
(AlaN$_3$) has been recently found to provide position-specific
information about the spectroscopy and dynamics of the modification
site.\cite{MS.lys:2021} The structure of the protein with the labelled
Ala residues is shown in Figure \ref{fig:lys} together with the
binding site lined by residues Leu84, Val87, Leu91, Leu99, Met102,
Val111, Ala112, Phe114, Ser117, Leu118, Leu121, Leu133,
Phe153. Following earlier work,\cite{MM.lys:2017} benzene was replaced
by cyano-benzene (PhCN) maintaining carbon atom positions. The WT
structure was used here to a) compare directly with earlier
results\cite{MS.lys:2021} and b) because PhCN has a comparatively
small binding free energy towards the WT protein ($\Delta G_{\rm bind}
= -0.5$ kcal/mol) which suggests that the interaction between the
ligand and the protein is weak.\cite{MM.lys:2017} For the L99A mutant
protein $\Delta G_{\rm bind} = -3.9$ kcal/mol for
PhCN\cite{MM.lys:2017} which compares with an experimentally
determined value of $\sim -3.5$ kcal/mol for iodobenzene from
isothermal titration calorimetry.\cite{liu:2009}\\

\noindent
For the -N$_3$ label a full-dimensional, accurate potential energy
surface (PES) calculated at the pair natural orbital based coupled
cluster
(PNO-LCCSD(T)-F12/aVTZ)\cite{lccsd-schwilk-2017,lccsdf12-schwilk-2017}
level and represented as a reproducing kernel Hilbert space
(RKHS)\cite{RKHS-Rabitz-1996,MM.rkhs:2017} is
available.\cite{MS.n3:2019} This energy function is suitable for
spectroscopic investigations and was combined with the CHARMM force
field\cite{charmmFF22} for the surrounding protein.\cite{MS.lys:2021}
MD simulations for the WT and all modified AlaN$_3$ labels were
carried out using an adapted version of the CHARMM
program\cite{Charmm-Brooks-2009} with an interface to perform the
simulations with the RKHS PES.\cite{MS.n3:2019} The protein is
solvated in explicit TIP3P water\cite{TIP3P-Jorgensen-1983} using a
cubic box of size $(78)^3$ \AA\/$^3$. First, all systems were
minimized followed by heating and equilibration. Next, 2 ns $NVT$
production simulations were carried out with and without the ligand
present in the active site for all 16 protein variants with Ala
replaced by AlaN$_3$. Bonds involving H-atoms were constrained using
the SHAKE\cite{SHAKE-Gunsteren-1997} algorithm and all nonbonded
interactions were evaluated with shifted interactions using a cutoff
of 14 \AA\/ and switched at 10 \AA\/.\cite{Steinbach1994} Snapshots
for analysis were recorded every 5 fs.\\

\begin{figure}[H]
\begin{center}
\includegraphics[width=0.9\textwidth]{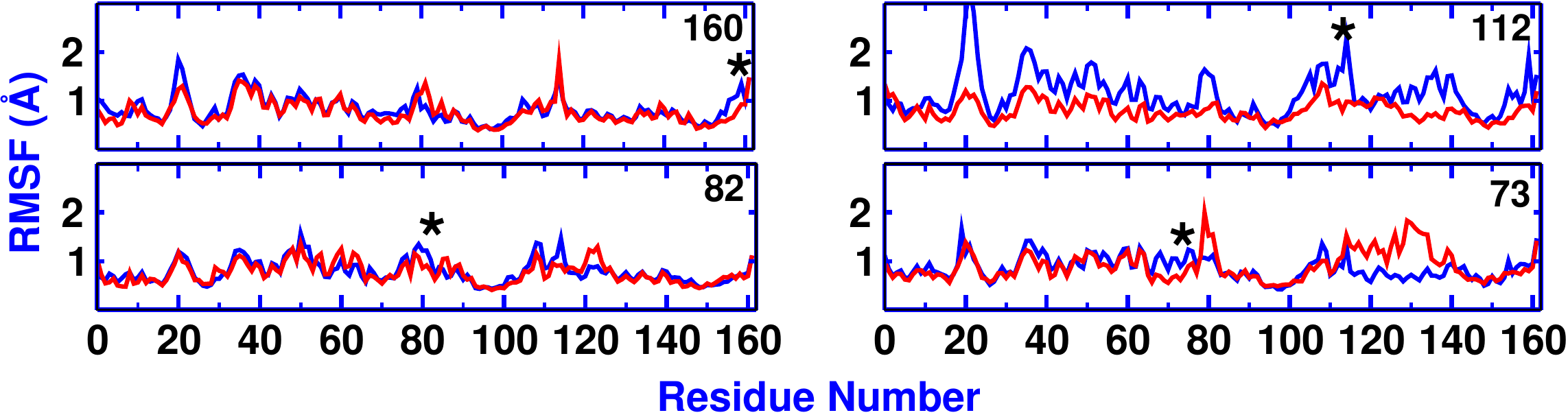}
\caption{RMSFs for the C$_{\alpha}$ atoms for ligand-free (blue) and
  ligand-bound (red) Lysozyme with N$_3^{-}$ attached to Ala73, Ala82,
  Ala112 and Ala160 residues. The label in each panel refers to the
  Ala residue number which carries the azide label and the
  corresponding position of residue is indicated with an asterisk
  above the RMSF trace.}
\label{fig:rmsf}
\end{center}
\end{figure}

\noindent
The effect of ligand binding on the overall flexibility of the
modified protein can be assessed from considering the root mean
squared fluctuation (RMSF) of the C$_{\alpha}$ atoms. Depending on the
position at which the -N$_3$ label is located, the changes in RMSF
range from insignificant (Ala82N$_3$ or Ala160N$_3$) to major
(Ala73N$_3$ or Ala112N$_3$), see Figures \ref{fig:rmsf} and S1.\\

\begin{figure}[H]
\begin{center}
\includegraphics[width=\textwidth]{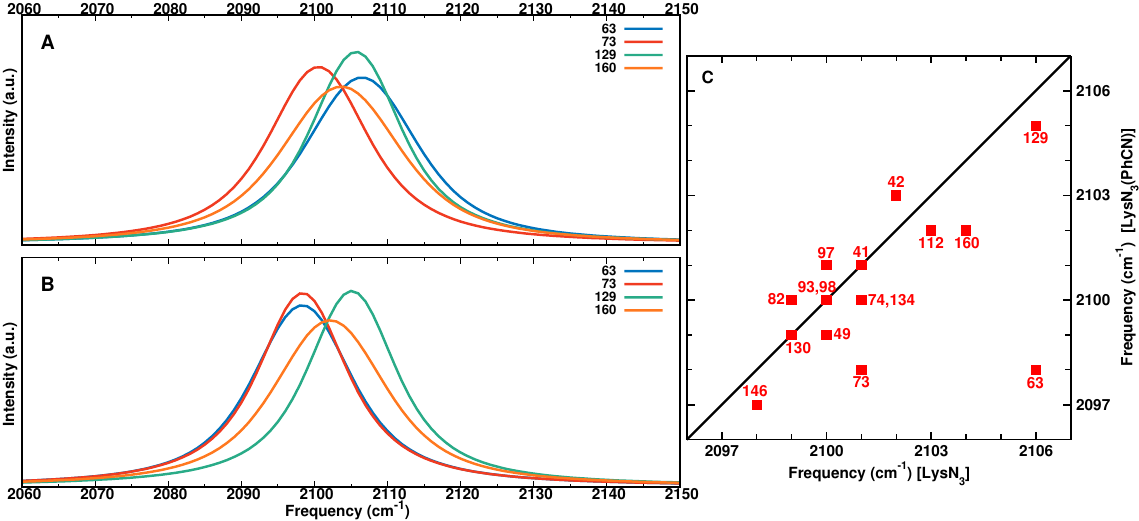}
\caption{1D IR spectra from INM for four AlaN$_3$ residues (Ala63,
  Ala73, Ala129, and Ala160) for ligand-free (panel A) and
  ligand-bound (panel B) Lysozyme. Panel C compares the maximum
  frequency of the 1D IR spectra for all modified Ala residues for
  ligand-free (along the $x-$acis) and ligand-bound (along the
  $y-$axis) N$_3$-labelled lysozyme.}
\label{fig:fmax}
\end{center}
\end{figure}

\noindent
Using instantaneous normal mode (INM)
analysis\cite{stratt:1994,MS.lys:2021,MS.n3:2019} the frequency
trajectory $\omega(t)$ of the asymmetric stretch vibration of the
-N$_3$ label was determined. Based on this, the 1D infrared spectra
corresponding to the azide asymmetric stretch vibration for each of
the 16 AlaN$_3$ residues was computed for the ligand-free and ligand-bound protein, see Figures \ref{fig:fmax} and S2. Direct comparison of the maximum position of the
infrared lineshape shows that for three N$_3$-modified alanine
residues (Ala41, Ala98, Ala130) the difference in the absorption
frequency is insignificant. For positions Ala63, Ala73 and Ala160 the
differences are 8, 3, and 2 cm$^{-1}$, respectively, whereas for the
other residues the change is within 1 cm$^{-1}$. Such frequency
changes can be measured with stat-of-the art
experiments\cite{thielges:2019} and their magnitude is also consistent
with previous simulations of the vibrational Stark effect for the -CN
probe in PhCN with red shifts of up to 3.5 cm$^{-1}$ in going from the
WT to the L99A and L99G mutants of T4-Lysozyme\cite{MM.lys:2017}
Similarly, the 1D and 2D infrared spectroscopy of -CO as the label for
insulin monomer and dimer found\cite{MS.insulin:2020} that the
relative shifts of the spectroscopic response was correctly described
whereas the absolute frequencies may differ by some 10 cm$^{-1}$. In a
very recent work such an approach found a splitting of 13 cm$^{-1}$,
compared with 25 cm$^{-1}$ from experiment, for the outer and central
-CO labels in cationic trialanine in water.\cite{MM.ala3:2021} Hence,
MD simulations together with instantaneous normal modes are a
meaningful approach to determine relative frequency shifts whereas
capturing absolute frequencies in such simulations requires slight
reparametrization of the underlying force field, e.g. through morphing
techniques.\cite{MM.morph:1999,JMB91morphing}\\

\noindent
The magnitude of frequency shifts found from the present simulations
is also comparable with --3 cm$^{-1}$ found from experiments of the
nitrile stretch in ligand IDD743 bound to WT vs. V47N mutant
hALR2\cite{Suydam.halr.sci.2006} or a $6$ cm$^{-1}$ blue shift of the
-CO vibrational frequency due to the binding of 19-NT to the Asp40Asn
mutant of the protein ketosteroid isomerase compared to the
WT.\cite{boxer.science:2014} Thus, differences of $\sim 1$ cm$^{-1}$
for the frequency of the reporter in different chemical environments
can be experimentally detected.\cite{thielges:2019}\\

\begin{figure}[H]
\begin{center}
\includegraphics[width=\textwidth]{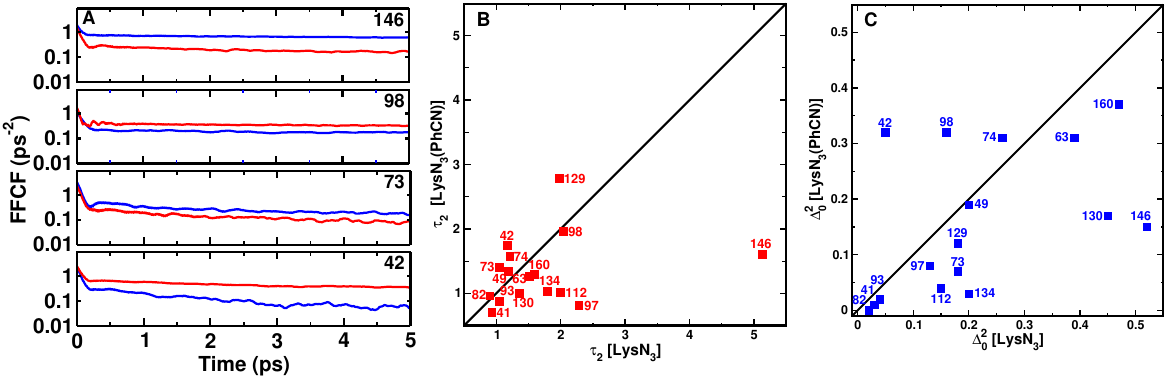}
\caption{Panel A: FFCFs with pronounced differences from correlating
  the INM frequencies for ligand-free and -bound Ala73, Ala146, Ala42,
  Ala98 in Lysozyme. The labels in each panel refer to the Ala residue
  which carries the azide label. Blue (ligand-free) and red
  (ligand-bound) traces are the fits to Eq. \ref{eq:ffcffit}. The
  $y-$axis is logarithmic. Panels B and C compare $\tau_2$ and
  $\Delta_0^2$ for ligand-bound and ligand-free Lysozyme,
  respectively.}
\label{fig:ffcf_corr}
\end{center}
\end{figure}

\noindent
From the frequency trajectories the frequency fluctuation correlation
function (FFCF) can be determined which contains valuable information
on relaxation time scales corresponding to the solvent dynamics around
the solute. The FFCFs are fit to an empirical expression
\begin{equation}
  \langle \delta \omega(t) \delta \omega(0) \rangle = a_{1}
  \cos(\gamma t) e^{-t/\tau_{1}} + \sum_{i=2}^{n} a_{i}
  e^{-t/\tau_{i}} + \Delta_0^2
\label{eq:ffcffit}
\end{equation}
which allows analytical integration to obtain the lineshape
function\cite{hynes:2004} using an automated curve fitting tool from
the SciPy library.\cite{2020SciPy-NMeth} As was found for the RMSFs
and 1D IR spectra, the FFCFs from the simulations with and without the
ligand bound to the protein can be very similar or differ appreciably,
see Figure \ref{fig:ffcf_corr}. The slow decay time, $\tau_2$, of the
-N$_3$ asymmetric stretch mode of the label is typically shorter for
the ligand-bound protein compared to that without PhCN, see Figure
\ref{fig:ffcf_corr}B, although exceptions exist. For Ala97N$_3$,
Ala112N$_3$, and Ala134N$_3$ the slow relaxation time $\tau_2$ is
faster by 75 \% up to a factor of $\sim 2.5$ and for Ala146N$_3$ the
slow time scale, $\tau_2$, differs by a factor of $\sim 3$ between
ligand-free ($\tau_2 = 5.13$ ps) and ligand-bound ($\tau_2 = 1.61$ ps)
Lysozyme. For the other Alanine residues the $\tau_2$ times between
ligand-free and ligand-bound lysozyme are similar. As an exception,
for Ala129N$_3$ the decay is slowed down by $\sim 50$ \% for
PhCN-bound lysozyme. All FFCFs without (Figure S3) and with (Figure S4)
the ligand bound together with the parameters of the empirical fit
(Table S1) are given in the supporting information.\\

\noindent
As a last feature of the FFCF it is found that the static component
$\Delta_0$ can differ appreciably between ligand-free and -bound
lysozyme. The static offset $\Delta_0$ is an experimental observable
and characterizes the structural heterogeneity around the modification
site. There are only four alanine residues for which the static offset
is similar (Ala41, Ala49, Ala82, and Ala93) for ligand-bound and
ligand-free lysozyme. For all others the differences range from 15 \%
to a factor of $\sim 3$. As an example, for Ala73N$_3$ the difference
for $\Delta_0^2$ between bound and ligand-free lysozyme is a factor of
$\sim 2.5$ ($\Delta_0^2 = 0.18$ vs. 0.07 ps$^{-2}$ or $\Delta_0 =
0.42$ vs. $\Delta_0 = 0.26$ ps$^{-1}$) and for Ala146N$_3$ they differ
by a factor of $\sim 3.5$ ($\Delta_0^2 = 0.52$ vs. 0.15 ps$^{-2}$;
i.e. $\Delta_0 = 0.72$ vs. $\Delta_0 = 0.39$ ps$^{-1}$). Thus, the
environmental dynamics around the spectroscopic label can be
sufficiently perturbed by binding of a ligand in the protein active
site to be reported directly as an experimentally accessible quantity
with typical errors\cite{thielges:2019} between 0.1 cm$^{-1}$ and 0.3
cm$^{-1}$ ($\sim 0.05$ ps$^{-1}$). Hence, the differences found from
the simulations are well outside the expected error bars from
experiment.\\

\noindent
Nonvanishing static components of the FFCF were also reported from
experiments. For trialanine (Ala)$_3$ a value of $\Delta_0 = 5$
cm$^{-1}$ was reported\cite{hamm:2002} compared with $\Delta_0 = 4.6$
cm$^{-1}$ from MD simulations (0.94 ps$^{-1}$ vs. 0.86 ps$^{-1}$) with
multipolar force fields.\cite{MM.ala3:2021} Similarly, CN$^{-}$ in
water features a nonvanishing tilt angle by $\tau = 10$
ps\cite{Kozinski07p5} with $\Delta_0 \sim 0.1$ ps$^{-1}$ $\sim 0.5$
cm$^{-1}$.\cite{MM.cn:2013} Finally, 2D IR experiments for
p-cyanophenylalanine bound to six distinct sites in a Src homology 3
domain reported static components ranging from $\Delta_0 = 1.0$ to 3.7
cm$^{-1}$ (corresponding to 0.19 ps$^{-1}$ to 0.70
ps$^{-1}$).\cite{thielges:2019}\\

\noindent
To determine in which way the dynamics of residues is affected upon
modification of the protein, dynamical cross-correlation
maps\cite{karplus:1991,ornstein1997} (DCCM) were calculated from the
trajectories using the Bio3D package.\cite{bio3d} Dynamic
cross-correlation matrices are based on the expression
\begin{equation}
 C_{ij} =  \langle \Delta r_{i}. \Delta r_{j}\rangle / (\langle \Delta r_{i}^{2} \langle \Delta r_{j}^{2}\rangle)^{1/2}
\label{eq:dccm}
\end{equation}
where $r_{i}$ and $r_{j}$ are the spatial C$_{\alpha}$ atom positions
of the respective $i$th and $j$th amino acids and $\Delta r_{i}$
corresponds to the displacement of the $i$th C$_{\alpha}$ from its
averaged position over the entire trajectory. DCCMs report on the
correlated and anticorrelated motions within a protein and difference
maps provide a global view of the positionally resolved differences in
the dynamics. In the following, only absolute values for $C_{ij}$ and
differences between them that are larger than 0.5 are reported. The
DCCMs are symmetrical about the diagonal and for clarity, positive
correlations (for DCCM) or positive differences in $C_{ij}$ (for
$\Delta$DCCM) are displayed in the lower right triangle and negative
values or differences in $C_{ij}$ are displayed in the upper left
triangle, respectively.\\

\begin{figure}[H]
\begin{center}
\includegraphics[width=1.0\textwidth]{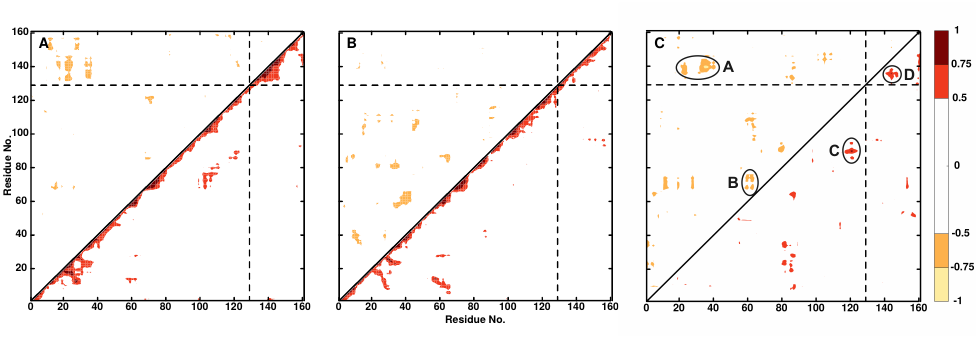}
\caption{DCCM for ligand-free (panel A), ligand-bound (panel B), and
  $\Delta$DCCM between ligand-free and -bound for
  Ala129N$_3$-PhCN. Positive correlations are in the lower right
  triangle, negative correlations in the upper left triangle. Only
  correlation coefficients with an absolute value greater than 0.5 are
  displayed.}
\label{fig:dccm129}
\end{center}
\end{figure}

\noindent
The DCCM for Lysozyme with Ala129N$_3$ ligand-free, ligand-bound and
the difference between the two is shown in Figure
\ref{fig:dccm129}. These maps reveal ligand-induced differences in the
correlated and anticorrelated motions with appreciable amplitudes, see
features A to D in Figure \ref{fig:dccm129}C. For ligand-free Lysozyme
there are pronounced couplings between residues $[130,147]$ and
$[20,25]$/$[32,37]$ in anticorrelated motions and residues $[68,80]$
and $[103,112]$ for correlated motions. As demonstrated in Figure
\ref{fig:dccm129}B, upon binding the ligand to Lysozyme the DCCM shows
different coupled residues compared to the ligand-free protein. As an
example, residues $[35,45]$ and $[55,68]$ are affected more for
anticorrelated motions while in the correlation ones, the coupling is
between residues $[5,15]$ and $[55,65]$. Note that these effects may
not be visible in the $\Delta$DCCM map as the magnitude of the
difference between the two systems may be below the threshold of 0.5
in the $\Delta C_{ij}$.\\

\noindent
In the difference map (Figure \ref{fig:dccm129}C) feature A indicates
the coupling between residues $[135,145]$ and $[20,25]$/$[30,42]$
whereas feature B refers to coupled residues $[65,75]$ and
$[58,65]$. Furthermore, feature C demonstrates prominent variations
between residues $[84,95]$ and $[117,125]$ while for feature D
residues $[129,140]$ and $[140,147]$ are strongly correlated. These
findings suggest that residues couple both locally (features B/D) and
through space (features A/C). It should also be pointed out that
residues involved in features A to C are among those with higher RMSF,
see Figure S1. Interestingly, the region around residue
Ala146 with larger differences $\Delta C_{ij}$ display correlated
dynamics with spatially close residues around residue Asp20 (white
licorice in Figure \ref{fig:lys}). On the other hand, the pronounced
differences in the RMSF of Ala129N$_3$ (see Figure S1)
for residues [42,57] do not show up in the $\Delta$DCCM map because
their $C_{ij}$ coefficients are below the threshold of 0.5.\\

\begin{figure}[H]
\begin{center}
\includegraphics[width=1.0\textwidth]{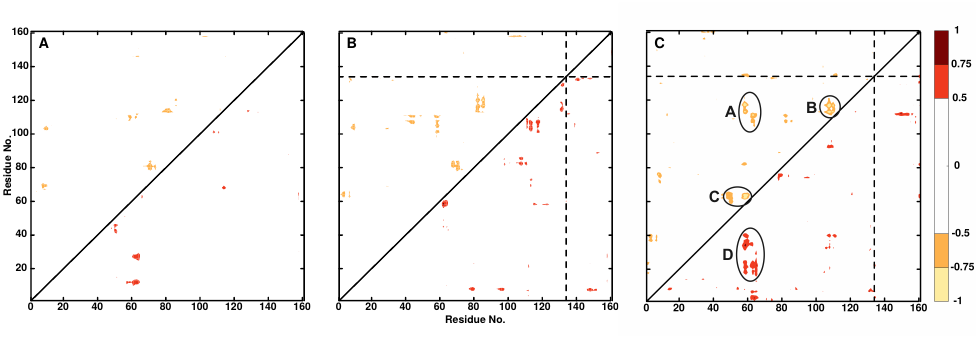}
\caption{Differences in DCCM maps ($\Delta$DCCM) between WT and
  WT-PhCN (panel A), WT and Ala134N$_3$ (panel B) and Ala134N$_3$ and
  Ala134N$_3$-PhCN (panel C). Positive correlations are in the lower
  right triangle, negative correlations in the upper left
  triangle. Only differences in correlation coefficients with an
  absolute value greater than 0.5 are displayed.}
\label{fig:ddccmwt_cn_134n3}
\end{center}
\end{figure}

\noindent
Difference $\Delta$DCCM maps between WT and ligand-bound (WT-PhCN)
Lysozyme are shown in Figure \ref{fig:ddccmwt_cn_134n3}A together with
the difference map between WT and azido modified lysozyme at position
134 (Ala134N$_3$), see Figure \ref{fig:ddccmwt_cn_134n3}B. With the
PhCN ligand bound to the protein the $\Delta$DCCM map compared with
that for the ligand-free protein is sparsely populated, see Figure
\ref{fig:ddccmwt_cn_134n3}A. This indicates that the conformational
dynamics of the two systems is similar. Contrary to that, more
differences in the dynamics between WT and Ala134N$_3$ arise as Figure
\ref{fig:ddccmwt_cn_134n3}B shows. Finally, the $\Delta$DCCM map
between ligand-free (Ala134N$_3$) and ligand-bound (Ala134N$_3$-PhCN)
labelled lysozyme at Ala134 shown in Figure
\ref{fig:ddccmwt_cn_134n3}C demonstrates that the ``contrast'' further
increases. The major difference in the conformational dynamics between
the ligand-free and ligand-bound protein arises for coupled residues
$[57,65]$ with $[105,120]$ (feature A), $[103,112]$ with $[112,122]$
(feature B), $[60,68]$ with $[47,52]$/$[55,62]$ (feature C), and
$[57,65]$ with $[17,25]$/$[32,42]$ (feature D). Interestingly, as
mentioned before for residue 129, residues involved in features A and
B are also among those with higher RMSF, see Figure S1. Additional $\Delta$DCCM plots for the remaining
AlaN$_3$ residues are shown in Figures S5 to
S17.\\

\noindent
In summary, the present work demonstrates that the 1D and 2D IR
spectroscopy of azide bound to alanine residues in WT Lysozyme
provides valuable site-specific and temporal information about ligand
binding of PhCN to the active site of WT lysozyme. Of particular note
is the increase in contrast between the ligand-free and the
ligand-bound protein when the azido-label is present, as demonstrated
for Ala134N$_3$. Furthermore, the static component $\Delta_0$ of the
FFCF, which is an experimentally accessible observable, shows
pronounced differences between the ligand-bound and ligand-free
protein and can serve as a useful indicator for ligand
binding. Changes in the maximum of the infrared absorbance are of the
order of one to several cm$^{-1}$ which is still detectable with
state-of-the-art experiments.\cite{thielges:2019} Given that the
-N$_3$ label can be introduced at multiple positions along the
polypeptide chain with specific spectroscopic signatures for each
variant of the system, it may even be possible to use the present
approach to refine existing structural models based on NMR
measurements\cite{becker:2018} or from more conventional
co-crystallization and X-ray structure determination efforts.\\

\section*{Acknowledgments}
The authors gratefully acknowledge financial support from the Swiss
National Science Foundation through grant 200021-117810 and to the
NCCR-MUST.\\

\section*{Supporting Information}
The supporting information contains Table S1 and Figures S1 to S17.

\section*{Data Availability Statement}
The data that support the findings of this study are available from
the corresponding author upon reasonable request.

\bibliography{lyso}

\providecommand{\latin}[1]{#1}
\makeatletter
\providecommand{\doi}
  {\begingroup\let\do\@makeother\dospecials
  \catcode`\{=1 \catcode`\}=2 \doi@aux}
\providecommand{\doi@aux}[1]{\endgroup\texttt{#1}}
\makeatother
\providecommand*\mcitethebibliography{\thebibliography}
\csname @ifundefined\endcsname{endmcitethebibliography}
  {\let\endmcitethebibliography\endthebibliography}{}
\begin{mcitethebibliography}{59}
\providecommand*\natexlab[1]{#1}
\providecommand*\mciteSetBstSublistMode[1]{}
\providecommand*\mciteSetBstMaxWidthForm[2]{}
\providecommand*\mciteBstWouldAddEndPuncttrue
  {\def\EndOfBibitem{\unskip.}}
\providecommand*\mciteBstWouldAddEndPunctfalse
  {\let\EndOfBibitem\relax}
\providecommand*\mciteSetBstMidEndSepPunct[3]{}
\providecommand*\mciteSetBstSublistLabelBeginEnd[3]{}
\providecommand*\EndOfBibitem{}
\mciteSetBstSublistMode{f}
\mciteSetBstMaxWidthForm{subitem}{(\alph{mcitesubitemcount})}
\mciteSetBstSublistLabelBeginEnd
  {\mcitemaxwidthsubitemform\space}
  {\relax}
  {\relax}

\bibitem[Plitzko \latin{et~al.}({2017})Plitzko, Schuler, and
  Selenko]{schuler:2017}
Plitzko,~J.~M.; Schuler,~B.; Selenko,~P. {Structural Biology outside the box -
  inside the cell}. \emph{Curr. Op. Struct. Biol.} \textbf{{2017}},
  \emph{{46}}, {110--121}\relax
\mciteBstWouldAddEndPuncttrue
\mciteSetBstMidEndSepPunct{\mcitedefaultmidpunct}
{\mcitedefaultendpunct}{\mcitedefaultseppunct}\relax
\EndOfBibitem
\bibitem[Guo and Zhou({2016})Guo, and Zhou]{zhou:2016}
Guo,~J.; Zhou,~H.-X. {Protein Allostery and Conformational Dynamics}.
  \emph{Chem. Rev.} \textbf{{2016}}, \emph{{116}}, {6503--6515}\relax
\mciteBstWouldAddEndPuncttrue
\mciteSetBstMidEndSepPunct{\mcitedefaultmidpunct}
{\mcitedefaultendpunct}{\mcitedefaultseppunct}\relax
\EndOfBibitem
\bibitem[Lu \latin{et~al.}({2019})Lu, He, Ni, and Zhang]{zhang:2019}
Lu,~S.; He,~X.; Ni,~D.; Zhang,~J. {Allosteric Modulator Discovery: From
  Serendipity to Structure-Based Design}. \emph{J. Med. Chem.} \textbf{{2019}},
  \emph{{62}}, {6405--6421}\relax
\mciteBstWouldAddEndPuncttrue
\mciteSetBstMidEndSepPunct{\mcitedefaultmidpunct}
{\mcitedefaultendpunct}{\mcitedefaultseppunct}\relax
\EndOfBibitem
\bibitem[Hamm and Zanni(2011)Hamm, and Zanni]{2DIRbook-Hamm-2011}
Hamm,~P.; Zanni,~M. \emph{{Concepts and Methods of 2D Infrared Spectroscopy}};
  Cambridge University Press: New York, 2011\relax
\mciteBstWouldAddEndPuncttrue
\mciteSetBstMidEndSepPunct{\mcitedefaultmidpunct}
{\mcitedefaultendpunct}{\mcitedefaultseppunct}\relax
\EndOfBibitem
\bibitem[Ma \latin{et~al.}({2015})Ma, Pazos, Zhang, Culik, and Gai]{gai:2015}
Ma,~J.; Pazos,~I.~M.; Zhang,~W.; Culik,~R.~M.; Gai,~F. {Site-Specific Infrared
  Probes of Proteins}. \emph{{Annu. Rev. Phys. Chem}} \textbf{{2015}},
  \emph{{66}}, {357--377}\relax
\mciteBstWouldAddEndPuncttrue
\mciteSetBstMidEndSepPunct{\mcitedefaultmidpunct}
{\mcitedefaultendpunct}{\mcitedefaultseppunct}\relax
\EndOfBibitem
\bibitem[Kim and Cho({2013})Kim, and Cho]{cho:2013}
Kim,~H.; Cho,~M. {Infrared Probes for Studying the Structure and Dynamics of
  Biomolecules}. \emph{{Chem. Rev.}} \textbf{{2013}}, \emph{{113}},
  {5817--5847}\relax
\mciteBstWouldAddEndPuncttrue
\mciteSetBstMidEndSepPunct{\mcitedefaultmidpunct}
{\mcitedefaultendpunct}{\mcitedefaultseppunct}\relax
\EndOfBibitem
\bibitem[Waegele \latin{et~al.}({2011})Waegele, Culik, and Gai]{gai:2011}
Waegele,~M.~M.; Culik,~R.~M.; Gai,~F. {Site-Specific Spectroscopic Reporters of
  the Local Electric Field, Hydration, Structure, and Dynamics of
  Biomolecules}. \emph{J. Phys. Chem. Lett.} \textbf{{2011}}, \emph{{2}},
  {2598--2609}\relax
\mciteBstWouldAddEndPuncttrue
\mciteSetBstMidEndSepPunct{\mcitedefaultmidpunct}
{\mcitedefaultendpunct}{\mcitedefaultseppunct}\relax
\EndOfBibitem
\bibitem[Koziol \latin{et~al.}({2015})Koziol, Johnson, Stucki-Buchli, Waldauer,
  and Hamm]{hamm.rev:2015}
Koziol,~K.~L.; Johnson,~P. J.~M.; Stucki-Buchli,~B.; Waldauer,~S.~A.; Hamm,~P.
  {Fast Infrared Spectroscopy of Protein Dynamics: Advancing Sensitivity and
  Selectivity}. \emph{Curr. Op. Struct. Biol.} \textbf{{2015}}, \emph{{34}},
  {1--6}\relax
\mciteBstWouldAddEndPuncttrue
\mciteSetBstMidEndSepPunct{\mcitedefaultmidpunct}
{\mcitedefaultendpunct}{\mcitedefaultseppunct}\relax
\EndOfBibitem
\bibitem[Levinson and Boxer({2014})Levinson, and Boxer]{boxer:2014}
Levinson,~N.~M.; Boxer,~S.~G. {A conserved water-mediated hydrogen bond network
  defines bosutinib's kinase selectivity}. \emph{{Nat. Chem. Biol.}}
  \textbf{{2014}}, \emph{{10}}, {127--132}\relax
\mciteBstWouldAddEndPuncttrue
\mciteSetBstMidEndSepPunct{\mcitedefaultmidpunct}
{\mcitedefaultendpunct}{\mcitedefaultseppunct}\relax
\EndOfBibitem
\bibitem[Layfield and Hammes-Schiffer({2013})Layfield, and
  Hammes-Schiffer]{hammes-schiffer:2013}
Layfield,~J.~P.; Hammes-Schiffer,~S. {Calculation of Vibrational Shifts of
  Nitrile Probes in the Active Site of Ketosteroid Isomerase upon Ligand
  Binding}. \emph{J. Am. Chem. Soc.} \textbf{{2013}}, \emph{{135}},
  {717--725}\relax
\mciteBstWouldAddEndPuncttrue
\mciteSetBstMidEndSepPunct{\mcitedefaultmidpunct}
{\mcitedefaultendpunct}{\mcitedefaultseppunct}\relax
\EndOfBibitem
\bibitem[Kuroda \latin{et~al.}({2013})Kuroda, Bauman, Challa, Patel, Troxler,
  Das, Arnold, and Hochstrasser]{hochstrasser:2013}
Kuroda,~D.~G.; Bauman,~J.~D.; Challa,~J.~R.; Patel,~D.; Troxler,~T.; Das,~K.;
  Arnold,~E.; Hochstrasser,~R.~M. {Snapshot of the equilibrium dynamics of a
  drug bound to HIV-1 reverse transcriptase}. \emph{{Nat. Chem.}}
  \textbf{{2013}}, \emph{{5}}, {174--181}\relax
\mciteBstWouldAddEndPuncttrue
\mciteSetBstMidEndSepPunct{\mcitedefaultmidpunct}
{\mcitedefaultendpunct}{\mcitedefaultseppunct}\relax
\EndOfBibitem
\bibitem[Liu \latin{et~al.}({2009})Liu, Strzalka, Tronin, Johansson, and
  Blasie]{blasie:2009}
Liu,~J.; Strzalka,~J.; Tronin,~A.; Johansson,~J.~S.; Blasie,~J.~K. {Mechanism
  of Interaction between the General Anesthetic Halothane and a Model Ion
  Channel Protein, II: Fluorescence and Vibrational Spectroscopy Using a
  Cyanophenylalanine Probe}. \emph{{Biophys. J.}} \textbf{{2009}}, \emph{{96}},
  {4176--4187}\relax
\mciteBstWouldAddEndPuncttrue
\mciteSetBstMidEndSepPunct{\mcitedefaultmidpunct}
{\mcitedefaultendpunct}{\mcitedefaultseppunct}\relax
\EndOfBibitem
\bibitem[Brewer \latin{et~al.}({2007})Brewer, Song, Raleigh, and
  Dyer]{dayer:2007}
Brewer,~S.~H.; Song,~B.; Raleigh,~D.~P.; Dyer,~R.~B. {Residue specific
  resolution of protein folding dynamics using isotope-edited infrared
  temperature jump spectroscopy}. \emph{{Biochem.}} \textbf{{2007}},
  \emph{{46}}, {3279--3285}\relax
\mciteBstWouldAddEndPuncttrue
\mciteSetBstMidEndSepPunct{\mcitedefaultmidpunct}
{\mcitedefaultendpunct}{\mcitedefaultseppunct}\relax
\EndOfBibitem
\bibitem[Middleton \latin{et~al.}({2012})Middleton, Marek, Cao, Chiu, Singh,
  Woys, de~Pablo, Raleigh, and Zanni]{zanni:2012}
Middleton,~C.~T.; Marek,~P.; Cao,~P.; Chiu,~C.-c.; Singh,~S.; Woys,~A.~M.;
  de~Pablo,~J.~J.; Raleigh,~D.~P.; Zanni,~M.~T. {Two-dimensional infrared
  spectroscopy reveals the complex behaviour of an amyloid fibril inhibitor}.
  \emph{{Nat. Chem.}} \textbf{{2012}}, \emph{{4}}, {355--360}\relax
\mciteBstWouldAddEndPuncttrue
\mciteSetBstMidEndSepPunct{\mcitedefaultmidpunct}
{\mcitedefaultendpunct}{\mcitedefaultseppunct}\relax
\EndOfBibitem
\bibitem[Ghosh \latin{et~al.}({2011})Ghosh, Qiu, DeGrado, and
  Hochstrasser]{hochstrasser:2011}
Ghosh,~A.; Qiu,~J.; DeGrado,~W.~F.; Hochstrasser,~R.~M. {Tidal surge in the M2
  proton channel, sensed by 2D IR spectroscopy}. \emph{Proc. Natl. Acad. Sci.}
  \textbf{{2011}}, \emph{{108}}, {6115--6120}\relax
\mciteBstWouldAddEndPuncttrue
\mciteSetBstMidEndSepPunct{\mcitedefaultmidpunct}
{\mcitedefaultendpunct}{\mcitedefaultseppunct}\relax
\EndOfBibitem
\bibitem[Remorino \latin{et~al.}({2011})Remorino, Korendovych, Wu, DeGrado, and
  Hochstrasser]{hochstrasser.science:2011}
Remorino,~A.; Korendovych,~I.~V.; Wu,~Y.; DeGrado,~W.~F.; Hochstrasser,~R.~M.
  {Residue-Specific Vibrational Echoes Yield 3D Structures of a Transmembrane
  Helix Dimer}. \emph{{Science}} \textbf{{2011}}, \emph{{332}},
  {1206--1209}\relax
\mciteBstWouldAddEndPuncttrue
\mciteSetBstMidEndSepPunct{\mcitedefaultmidpunct}
{\mcitedefaultendpunct}{\mcitedefaultseppunct}\relax
\EndOfBibitem
\bibitem[van Wilderen \latin{et~al.}({2014})van Wilderen, Kern-Michler,
  Mueller-Werkmeister, and Bredenbeck]{bredenbeck:2014}
van Wilderen,~L. J. G.~W.; Kern-Michler,~D.; Mueller-Werkmeister,~H.~M.;
  Bredenbeck,~J. {Vibrational dynamics and solvatochromism of the label SCN in
  various solvents and hemoglobin by time dependent IR and 2D-IR spectroscopy}.
  \emph{Phys. Chem. Chem. Phys.} \textbf{{2014}}, \emph{{16}},
  {19643--19653}\relax
\mciteBstWouldAddEndPuncttrue
\mciteSetBstMidEndSepPunct{\mcitedefaultmidpunct}
{\mcitedefaultendpunct}{\mcitedefaultseppunct}\relax
\EndOfBibitem
\bibitem[Lee \latin{et~al.}({2018})Lee, Kossowska, Lim, Kim, Han, Kwak, and
  Cho]{cho:2018}
Lee,~G.; Kossowska,~D.; Lim,~J.; Kim,~S.; Han,~H.; Kwak,~K.; Cho,~M. {Cyanamide
  as an Infrared Reporter: Comparison of Vibrational Properties between
  Nitriles Bonded to N and C Atoms}. \emph{J. Phys. Chem. B} \textbf{{2018}},
  \emph{{122}}, {4035--4044}\relax
\mciteBstWouldAddEndPuncttrue
\mciteSetBstMidEndSepPunct{\mcitedefaultmidpunct}
{\mcitedefaultendpunct}{\mcitedefaultseppunct}\relax
\EndOfBibitem
\bibitem[Kozinski \latin{et~al.}({2008})Kozinski, Garrett-Roe, and
  Hamm]{hamm:2008}
Kozinski,~M.; Garrett-Roe,~S.; Hamm,~P. {2D-IR spectroscopy of the sulfhydryl
  band of cysteines in the hydrophobic core of proteins}. \emph{J. Phys. Chem.
  B} \textbf{{2008}}, \emph{{112}}, {7645--7650}\relax
\mciteBstWouldAddEndPuncttrue
\mciteSetBstMidEndSepPunct{\mcitedefaultmidpunct}
{\mcitedefaultendpunct}{\mcitedefaultseppunct}\relax
\EndOfBibitem
\bibitem[Zimmermann \latin{et~al.}({2011})Zimmermann, Thielges, Yu, Dawson, and
  Romesberg]{romesberg:2011}
Zimmermann,~J.; Thielges,~M.~C.; Yu,~W.; Dawson,~P.~E.; Romesberg,~F.~E.
  {Carbon-Deuterium Bonds as Site-Specific and Nonperturbative Probes for
  Time-Resolved Studies of Protein Dynamics and Folding}. \emph{J. Phys. Chem.
  Lett.} \textbf{{2011}}, \emph{{2}}, {412--416}\relax
\mciteBstWouldAddEndPuncttrue
\mciteSetBstMidEndSepPunct{\mcitedefaultmidpunct}
{\mcitedefaultendpunct}{\mcitedefaultseppunct}\relax
\EndOfBibitem
\bibitem[Woys \latin{et~al.}({2013})Woys, Mukherjee, Skoff, Moran, and
  Zanni]{zanni:2013}
Woys,~A.~M.; Mukherjee,~S.~S.; Skoff,~D.~R.; Moran,~S.~D.; Zanni,~M.~T. {A
  Strongly Absorbing Class of Non-Natural Labels for Probing Protein
  Electrostatics and Solvation with FTIR and 2D IR Spectroscopies}. \emph{J.
  Phys. Chem. B} \textbf{{2013}}, \emph{{117}}, {5009--5018}\relax
\mciteBstWouldAddEndPuncttrue
\mciteSetBstMidEndSepPunct{\mcitedefaultmidpunct}
{\mcitedefaultendpunct}{\mcitedefaultseppunct}\relax
\EndOfBibitem
\bibitem[Ramos \latin{et~al.}({2019})Ramos, Horness, Collins, Haak, and
  Thielges]{thielges:2019}
Ramos,~S.; Horness,~R.~E.; Collins,~J.~A.; Haak,~D.; Thielges,~M.~C.
  {Site-specific 2D IR spectroscopy: a general approach for the
  characterization of protein dynamics with high spatial and temporal
  resolution}. \emph{Phys. Chem. Chem. Phys.} \textbf{{2019}}, \emph{{21}},
  {780--788}\relax
\mciteBstWouldAddEndPuncttrue
\mciteSetBstMidEndSepPunct{\mcitedefaultmidpunct}
{\mcitedefaultendpunct}{\mcitedefaultseppunct}\relax
\EndOfBibitem
\bibitem[Bloem \latin{et~al.}({2012})Bloem, Koziol, Waldauer, Buchli, Walser,
  Samatanga, Jelesarov, and Hamm]{hamm:2012}
Bloem,~R.; Koziol,~K.; Waldauer,~S.~A.; Buchli,~B.; Walser,~R.; Samatanga,~B.;
  Jelesarov,~I.; Hamm,~P. {Ligand Binding Studied by 2D IR Spectroscopy Using
  the Azidohomoalanine Label}. \emph{J. Phys. Chem. B} \textbf{{2012}},
  \emph{{116}}, {13705--13712}\relax
\mciteBstWouldAddEndPuncttrue
\mciteSetBstMidEndSepPunct{\mcitedefaultmidpunct}
{\mcitedefaultendpunct}{\mcitedefaultseppunct}\relax
\EndOfBibitem
\bibitem[Johnson \latin{et~al.}({2017})Johnson, Koziol, and
  Hamm]{hamm.rev:2017}
Johnson,~P. J.~M.; Koziol,~K.~L.; Hamm,~P. {Quantifying Biomolecular
  Recognition with Site-Specific 2D Infrared Probes}. \emph{J. Phys. Chem.
  Lett.} \textbf{{2017}}, \emph{{8}}, {2280--2284}\relax
\mciteBstWouldAddEndPuncttrue
\mciteSetBstMidEndSepPunct{\mcitedefaultmidpunct}
{\mcitedefaultendpunct}{\mcitedefaultseppunct}\relax
\EndOfBibitem
\bibitem[Salehi and Meuwly({2021})Salehi, and Meuwly]{MS.lys:2021}
Salehi,~S.~M.; Meuwly,~M. {Site-selective dynamics of azidolysozyme}.
  \emph{{JCP}} \textbf{{2021}}, \emph{{154}}\relax
\mciteBstWouldAddEndPuncttrue
\mciteSetBstMidEndSepPunct{\mcitedefaultmidpunct}
{\mcitedefaultendpunct}{\mcitedefaultseppunct}\relax
\EndOfBibitem
\bibitem[Kiick \latin{et~al.}({2002})Kiick, Saxon, Tirrell, and
  Bertozzi]{bertozzi:2002}
Kiick,~K.; Saxon,~E.; Tirrell,~D.; Bertozzi,~C. {Incorporation of azides into
  recombinant proteins for chemoselective modification by the Staudinger
  ligation}. \emph{Proc. Natl. Acad. Sci.} \textbf{{2002}}, \emph{{99}},
  {19--24}\relax
\mciteBstWouldAddEndPuncttrue
\mciteSetBstMidEndSepPunct{\mcitedefaultmidpunct}
{\mcitedefaultendpunct}{\mcitedefaultseppunct}\relax
\EndOfBibitem
\bibitem[Zanobini \latin{et~al.}({2018})Zanobini, Bozovic, Jankovic, Koziol,
  Johnson, Hamm, Gulzar, Wolf, and Stock]{stock:2018}
Zanobini,~C.; Bozovic,~O.; Jankovic,~B.; Koziol,~K.~L.; Johnson,~P. J.~M.;
  Hamm,~P.; Gulzar,~A.; Wolf,~S.; Stock,~G. {Azidohomoalanine: A Minimally
  Invasive, Versatile, and Sensitive Infrared Label in Proteins To Study Ligand
  Binding}. \emph{J. Phys. Chem. B} \textbf{{2018}}, \emph{{122}},
  {10118--10125}\relax
\mciteBstWouldAddEndPuncttrue
\mciteSetBstMidEndSepPunct{\mcitedefaultmidpunct}
{\mcitedefaultendpunct}{\mcitedefaultseppunct}\relax
\EndOfBibitem
\bibitem[Wolfshorndl \latin{et~al.}({2012})Wolfshorndl, Baskin, Dhawan, and
  Londergan]{londergan:2012}
Wolfshorndl,~M.~P.; Baskin,~R.; Dhawan,~I.; Londergan,~C.~H. {Covalently Bound
  Azido Groups Are Very Specific Water Sensors, Even in Hydrogen-Bonding
  Environments}. \emph{J. Phys. Chem. B} \textbf{{2012}}, \emph{{116}},
  {1172--1179}\relax
\mciteBstWouldAddEndPuncttrue
\mciteSetBstMidEndSepPunct{\mcitedefaultmidpunct}
{\mcitedefaultendpunct}{\mcitedefaultseppunct}\relax
\EndOfBibitem
\bibitem[Oh \latin{et~al.}({2008})Oh, Lee, Joo, Han, and Cho]{cho:2008}
Oh,~K.-I.; Lee,~J.-H.; Joo,~C.; Han,~H.; Cho,~M. {beta-Azidoalanine as an IR
  probe: Application to amyloid A beta(16-22) aggregation}. \emph{J. Phys.
  Chem. B} \textbf{{2008}}, \emph{{112}}, {10352--10357}\relax
\mciteBstWouldAddEndPuncttrue
\mciteSetBstMidEndSepPunct{\mcitedefaultmidpunct}
{\mcitedefaultendpunct}{\mcitedefaultseppunct}\relax
\EndOfBibitem
\bibitem[Le~Sueur \latin{et~al.}({2016})Le~Sueur, Schaugaard, Baik, and
  Thielges]{thielges:2016}
Le~Sueur,~A.~L.; Schaugaard,~R.~N.; Baik,~M.-H.; Thielges,~M.~C. {Methionine
  Ligand Interaction in a Blue Copper Protein Characterized by Site-Selective
  Infrared Spectroscopy}. \emph{J. Am. Chem. Soc.} \textbf{{2016}},
  \emph{{138}}, {7187--7193}\relax
\mciteBstWouldAddEndPuncttrue
\mciteSetBstMidEndSepPunct{\mcitedefaultmidpunct}
{\mcitedefaultendpunct}{\mcitedefaultseppunct}\relax
\EndOfBibitem
\bibitem[Suydam \latin{et~al.}(2006)Suydam, Snow, Pande, and
  Boxer]{Suydam.halr.sci.2006}
Suydam,~I.~T.; Snow,~C.~D.; Pande,~V.~S.; Boxer,~S.~G. Electric Fields at the
  Active Site of an Enzyme : Direct Comparison of Experiment with Theory.
  \emph{Science} \textbf{2006}, \emph{313}, 200--204\relax
\mciteBstWouldAddEndPuncttrue
\mciteSetBstMidEndSepPunct{\mcitedefaultmidpunct}
{\mcitedefaultendpunct}{\mcitedefaultseppunct}\relax
\EndOfBibitem
\bibitem[Mondal and Meuwly(2017)Mondal, and Meuwly]{MM.lys:2017}
Mondal,~P.; Meuwly,~M. Vibrational Stark Spectroscopy for Assessing
  Ligand-Binding Strengths in a Protein. \emph{Phys. Chem. Chem. Phys.}
  \textbf{2017}, \emph{19}, 16131--16143\relax
\mciteBstWouldAddEndPuncttrue
\mciteSetBstMidEndSepPunct{\mcitedefaultmidpunct}
{\mcitedefaultendpunct}{\mcitedefaultseppunct}\relax
\EndOfBibitem
\bibitem[Liu \latin{et~al.}(2009)Liu, Baase, and Matthews]{liu:2009}
Liu,~L.; Baase,~W.~A.; Matthews,~B.~W. Halogenated benzenes bound within a
  non-polar cavity in T4 lysozyme provide examples of I⋯ S and I⋯ Se
  halogen-bonding. \emph{J. Mol. Biol.} \textbf{2009}, \emph{385},
  595--605\relax
\mciteBstWouldAddEndPuncttrue
\mciteSetBstMidEndSepPunct{\mcitedefaultmidpunct}
{\mcitedefaultendpunct}{\mcitedefaultseppunct}\relax
\EndOfBibitem
\bibitem[Schwilk \latin{et~al.}({2017})Schwilk, Ma, Koeppl, and
  Werner]{lccsd-schwilk-2017}
Schwilk,~M.; Ma,~Q.; Koeppl,~C.; Werner,~H.-J. {Scalable Electron Correlation
  Methods. 3. Efficient and Accurate Parallel Local Coupled Cluster with Pair
  Natural Orbitals (PNO-LCCSD)}. \emph{J. Chem. Theo. Comp.} \textbf{{2017}},
  \emph{{13}}, {3650--3675}\relax
\mciteBstWouldAddEndPuncttrue
\mciteSetBstMidEndSepPunct{\mcitedefaultmidpunct}
{\mcitedefaultendpunct}{\mcitedefaultseppunct}\relax
\EndOfBibitem
\bibitem[Ma \latin{et~al.}({2018})Ma, Schwilk, Koeppl, and
  Werner]{lccsdf12-schwilk-2017}
Ma,~Q.; Schwilk,~M.; Koeppl,~C.; Werner,~H.-J. {Scalable Electron Correlation
  Methods. 4. Parallel Explicitly Correlated Local Coupled Cluster with Pair
  Natural Orbitals (PNO-LCCSD-F12) (vol 13, pg 4871, 2017)}. \emph{J. Chem.
  Theo. Comp.} \textbf{{2018}}, \emph{{14}}, {6750}\relax
\mciteBstWouldAddEndPuncttrue
\mciteSetBstMidEndSepPunct{\mcitedefaultmidpunct}
{\mcitedefaultendpunct}{\mcitedefaultseppunct}\relax
\EndOfBibitem
\bibitem[Ho and Rabitz(1996)Ho, and Rabitz]{RKHS-Rabitz-1996}
Ho,~T.-S.; Rabitz,~R. A General Method for Constructing Multidimensional
  Molecular Potential Energy Surfaces from ab Initio Calculations. \emph{J.
  Chem. Phys.} \textbf{1996}, \emph{104}, 2584--2597\relax
\mciteBstWouldAddEndPuncttrue
\mciteSetBstMidEndSepPunct{\mcitedefaultmidpunct}
{\mcitedefaultendpunct}{\mcitedefaultseppunct}\relax
\EndOfBibitem
\bibitem[Unke and Meuwly(2017)Unke, and Meuwly]{MM.rkhs:2017}
Unke,~O.~T.; Meuwly,~M. Toolkit for the Construction of Reproducing
  Kernel-Based Representations of Data: Application to Multidimensional
  Potential Energy Surfaces. \emph{J. Chem. Inf. Model.} \textbf{2017},
  \emph{57}, 1923--1931\relax
\mciteBstWouldAddEndPuncttrue
\mciteSetBstMidEndSepPunct{\mcitedefaultmidpunct}
{\mcitedefaultendpunct}{\mcitedefaultseppunct}\relax
\EndOfBibitem
\bibitem[Salehi \latin{et~al.}({2019})Salehi, Koner, and Meuwly]{MS.n3:2019}
Salehi,~S.~M.; Koner,~D.; Meuwly,~M. {Vibrational Spectroscopy of N$_{3}^{-}$
  in the Gas and Condensed Phase}. \emph{J. Phys. Chem. B} \textbf{{2019}},
  \emph{{123}}, {3282--3290}\relax
\mciteBstWouldAddEndPuncttrue
\mciteSetBstMidEndSepPunct{\mcitedefaultmidpunct}
{\mcitedefaultendpunct}{\mcitedefaultseppunct}\relax
\EndOfBibitem
\bibitem[MacKerell \latin{et~al.}(1998)MacKerell, Bashford, Bellott, Dunbrack,
  Evanseck, Field, Fischer, Gao, Guo, Ha, Joseph-McCarthy, Kuchnir, Kuczera,
  Lau, Mattos, Michnick, Ngo, Nguyen, Prodhom, Reiher, Roux, Schlenkrich,
  Smith, Stote, Straub, Watanabe, Wirkiewicz-Kuczera, Yin, and
  Karplus]{charmmFF22}
MacKerell,~A.~D.; Bashford,~D.; Bellott,~M.; Dunbrack,~R.~L.; Evanseck,~J.~D.;
  Field,~M.~J.; Fischer,~S.; Gao,~J.; Guo,~H.; Ha,~S. \latin{et~al.}  All-atom
  Empirical Potential for Molecular Modeling and Dynamics Studies of Proteins.
  \emph{J. Phys. Chem. B} \textbf{1998}, \emph{102}, 3586--3616\relax
\mciteBstWouldAddEndPuncttrue
\mciteSetBstMidEndSepPunct{\mcitedefaultmidpunct}
{\mcitedefaultendpunct}{\mcitedefaultseppunct}\relax
\EndOfBibitem
\bibitem[Brooks \latin{et~al.}(2009)Brooks, Brooks~III, MacKerell~Jr., Nilsson,
  Petrella, Roux, Won, Archontis, Bartels, Boresch, Caflisch, Caves, Cui,
  Dinner, Feig, Fischer, Gao, Hodoscek, Im, Kuczera, Lazaridis, Ma,
  Ovchinnikov, Paci, Pastor, Post, Schaefer, Tidor, Venable, Woodcock, Wu,
  Yang, York, and Karplus]{Charmm-Brooks-2009}
Brooks,~B.~R.; Brooks~III,~C.~L.; MacKerell~Jr.,~A.~D.; Nilsson,~L.;
  Petrella,~R.~J.; Roux,~B.; Won,~Y.; Archontis,~G.; Bartels,~C.; Boresch,~S.
  \latin{et~al.}  CHARMM: The Biomolecular Simulation Program. \emph{J. Comp.
  Chem.} \textbf{2009}, \emph{30}, 1545--1614\relax
\mciteBstWouldAddEndPuncttrue
\mciteSetBstMidEndSepPunct{\mcitedefaultmidpunct}
{\mcitedefaultendpunct}{\mcitedefaultseppunct}\relax
\EndOfBibitem
\bibitem[Jorgensen \latin{et~al.}(1983)Jorgensen, Chandrasekhar, Madura, Impey,
  and Klein]{TIP3P-Jorgensen-1983}
Jorgensen,~W.~L.; Chandrasekhar,~J.; Madura,~J.~D.; Impey,~R.~W.; Klein,~M.~L.
  Comparison of Simple Potential Functions for Simulating Liquid Water.
  \emph{J. Chem. Phys.} \textbf{1983}, \emph{79}, 926--935\relax
\mciteBstWouldAddEndPuncttrue
\mciteSetBstMidEndSepPunct{\mcitedefaultmidpunct}
{\mcitedefaultendpunct}{\mcitedefaultseppunct}\relax
\EndOfBibitem
\bibitem[Gunsteren and Berendsen(1997)Gunsteren, and
  Berendsen]{SHAKE-Gunsteren-1997}
Gunsteren,~W.~V.; Berendsen,~H. Algorithms for Macromolecular Dynamics and
  Constraint Dynamics. \emph{Mol. Phys.} \textbf{1997}, \emph{34},
  1311--1327\relax
\mciteBstWouldAddEndPuncttrue
\mciteSetBstMidEndSepPunct{\mcitedefaultmidpunct}
{\mcitedefaultendpunct}{\mcitedefaultseppunct}\relax
\EndOfBibitem
\bibitem[Steinbach and Brooks(1994)Steinbach, and Brooks]{Steinbach1994}
Steinbach,~P.~J.; Brooks,~B.~R. New Spherical-Cutoff Methods for Long-Range
  Forces in Macromolecular Simulation. \emph{J. Comput. Chem.} \textbf{1994},
  \emph{15}, 667--683\relax
\mciteBstWouldAddEndPuncttrue
\mciteSetBstMidEndSepPunct{\mcitedefaultmidpunct}
{\mcitedefaultendpunct}{\mcitedefaultseppunct}\relax
\EndOfBibitem
\bibitem[Cho \latin{et~al.}({1994})Cho, Fleming, Saito, Ohmine, and
  Stratt]{stratt:1994}
Cho,~M.; Fleming,~G.; Saito,~S.; Ohmine,~I.; Stratt,~R. {Instantaneous Normal
  Mode Analysis of Liquid Water}. \emph{J. Chem. Phys.} \textbf{{1994}},
  \emph{{100}}, {6672--6683}\relax
\mciteBstWouldAddEndPuncttrue
\mciteSetBstMidEndSepPunct{\mcitedefaultmidpunct}
{\mcitedefaultendpunct}{\mcitedefaultseppunct}\relax
\EndOfBibitem
\bibitem[Salehi \latin{et~al.}({2020})Salehi, Koner, and
  Meuwly]{MS.insulin:2020}
Salehi,~S.~M.; Koner,~D.; Meuwly,~M. {Dynamics and Infrared Spectroscopy of
  Monomeric and Dimeric Wild Type and Mutant Insulin}. \emph{J. Phys. Chem. B}
  \textbf{{2020}}, \emph{{124}}, {11882--11894}\relax
\mciteBstWouldAddEndPuncttrue
\mciteSetBstMidEndSepPunct{\mcitedefaultmidpunct}
{\mcitedefaultendpunct}{\mcitedefaultseppunct}\relax
\EndOfBibitem
\bibitem[Mondal \latin{et~al.}(2021)Mondal, Cazade, Das, Bereau, and
  Meuwly]{MM.ala3:2021}
Mondal,~P.; Cazade,~P.-A.; Das,~A.~K.; Bereau,~T.; Meuwly,~M. Multipolar Force
  Fields for Amide-I Spectroscopy from Conformational Dynamics of the
  Alanine-Trimer. \emph{arXiv preprint arXiv:2106.10142, in print in J. Phys.
  Chem. B} \textbf{2021}, \relax
\mciteBstWouldAddEndPunctfalse
\mciteSetBstMidEndSepPunct{\mcitedefaultmidpunct}
{}{\mcitedefaultseppunct}\relax
\EndOfBibitem
\bibitem[Meuwly and Hutson(1999)Meuwly, and Hutson]{MM.morph:1999}
Meuwly,~M.; Hutson,~J.~M. Morphing ab Initio potentials: A systematic study of
  Ne--HF. \emph{J. Chem. Phys.} \textbf{1999}, \emph{110}, 8338--8347\relax
\mciteBstWouldAddEndPuncttrue
\mciteSetBstMidEndSepPunct{\mcitedefaultmidpunct}
{\mcitedefaultendpunct}{\mcitedefaultseppunct}\relax
\EndOfBibitem
\bibitem[Bowman and Gazdy(1991)Bowman, and Gazdy]{JMB91morphing}
Bowman,~J.~M.; Gazdy,~B. A simple method to adjust potential energy surfaces:
  Application to HCO. \emph{J. Chem. Phys.} \textbf{1991}, \emph{94},
  816--817\relax
\mciteBstWouldAddEndPuncttrue
\mciteSetBstMidEndSepPunct{\mcitedefaultmidpunct}
{\mcitedefaultendpunct}{\mcitedefaultseppunct}\relax
\EndOfBibitem
\bibitem[Fried \latin{et~al.}({2014})Fried, Bagchi, and
  Boxer]{boxer.science:2014}
Fried,~S.~D.; Bagchi,~S.; Boxer,~S.~G. {Extreme electric fields power catalysis
  in the active site of ketosteroid isomerase}. \emph{{Science}}
  \textbf{{2014}}, \emph{{346}}, {1510--1514}\relax
\mciteBstWouldAddEndPuncttrue
\mciteSetBstMidEndSepPunct{\mcitedefaultmidpunct}
{\mcitedefaultendpunct}{\mcitedefaultseppunct}\relax
\EndOfBibitem
\bibitem[Moller \latin{et~al.}({2004})Moller, Rey, and Hynes]{hynes:2004}
Moller,~K.; Rey,~R.; Hynes,~J. {Hydrogen Bond Dynamics in Water and Ultrafast
  Infrared Spectroscopy: A Theoretical Study}. \emph{J. Phys. Chem. A}
  \textbf{{2004}}, \emph{{108}}, {1275--1289}\relax
\mciteBstWouldAddEndPuncttrue
\mciteSetBstMidEndSepPunct{\mcitedefaultmidpunct}
{\mcitedefaultendpunct}{\mcitedefaultseppunct}\relax
\EndOfBibitem
\bibitem[{Virtanen} \latin{et~al.}(2020){Virtanen}, {Gommers}, {Oliphant},
  {Haberland}, {Reddy}, {Cournapeau}, {Burovski}, {Peterson}, {Weckesser},
  {Bright}, {van der Walt}, {Brett}, {Wilson}, {Jarrod Millman}, {Mayorov},
  {Nelson}, {Jones}, {Kern}, {Larson}, {Carey}, {Polat}, {Feng}, {Moore}, {Vand
  erPlas}, {Laxalde}, {Perktold}, {Cimrman}, {Henriksen}, {Quintero}, {Harris},
  {Archibald}, {Ribeiro}, {Pedregosa}, {van Mulbregt}, and
  {Contributors}]{2020SciPy-NMeth}
{Virtanen},~P.; {Gommers},~R.; {Oliphant},~T.~E.; {Haberland},~M.; {Reddy},~T.;
  {Cournapeau},~D.; {Burovski},~E.; {Peterson},~P.; {Weckesser},~W.;
  {Bright},~J. \latin{et~al.}  {SciPy 1.0: Fundamental Algorithms for
  Scientific Computing in Python}. \emph{Nat. Methods} \textbf{2020},
  \emph{17}, 261--272\relax
\mciteBstWouldAddEndPuncttrue
\mciteSetBstMidEndSepPunct{\mcitedefaultmidpunct}
{\mcitedefaultendpunct}{\mcitedefaultseppunct}\relax
\EndOfBibitem
\bibitem[Woutersen \latin{et~al.}(2002)Woutersen, Pfister, Hamm, Mu, Kosov, and
  Stock]{hamm:2002}
Woutersen,~S.; Pfister,~R.; Hamm,~P.; Mu,~Y.; Kosov,~D.~S.; Stock,~G. Peptide
  conformational heterogeneity revealed from nonlinear vibrational spectroscopy
  and molecular-dynamics simulations. \emph{J. Chem. Phys.} \textbf{2002},
  \emph{117}, 6833--6840\relax
\mciteBstWouldAddEndPuncttrue
\mciteSetBstMidEndSepPunct{\mcitedefaultmidpunct}
{\mcitedefaultendpunct}{\mcitedefaultseppunct}\relax
\EndOfBibitem
\bibitem[Kozi\'{n}ski \latin{et~al.}(2007)Kozi\'{n}ski, Garrett-Roe, and
  Hamm]{Kozinski07p5}
Kozi\'{n}ski,~M.; Garrett-Roe,~S.; Hamm,~P. Vibrational Spectral Diffusion of
  CN$^-$ in Water. \emph{Chem. Phys.} \textbf{2007}, \emph{341}, 5--10\relax
\mciteBstWouldAddEndPuncttrue
\mciteSetBstMidEndSepPunct{\mcitedefaultmidpunct}
{\mcitedefaultendpunct}{\mcitedefaultseppunct}\relax
\EndOfBibitem
\bibitem[Lee \latin{et~al.}(2013)Lee, Carr, G\"{o}llner, Hamm, and
  Meuwly]{MM.cn:2013}
Lee,~M.~W.; Carr,~J.~K.; G\"{o}llner,~M.; Hamm,~P.; Meuwly,~M. 2D IR Spectra of
  Cyanide in Water Investigated by Molecular Dynamics Simulations. \emph{J.
  Chem. Phys.} \textbf{2013}, \emph{139}, 054506\relax
\mciteBstWouldAddEndPuncttrue
\mciteSetBstMidEndSepPunct{\mcitedefaultmidpunct}
{\mcitedefaultendpunct}{\mcitedefaultseppunct}\relax
\EndOfBibitem
\bibitem[Ichiye and Karplus({1991})Ichiye, and Karplus]{karplus:1991}
Ichiye,~T.; Karplus,~M. {Collective Motions in Proteins - A Covariance Analysis
  of Atomic Fluctuation in Molecular-Dynamics and Normal Mode Simulations}.
  \emph{{Protein Struct. Funct. Genet.}} \textbf{{1991}}, \emph{{11}},
  {205--217}\relax
\mciteBstWouldAddEndPuncttrue
\mciteSetBstMidEndSepPunct{\mcitedefaultmidpunct}
{\mcitedefaultendpunct}{\mcitedefaultseppunct}\relax
\EndOfBibitem
\bibitem[Arnold and Ornstein({1997})Arnold, and Ornstein]{ornstein1997}
Arnold,~G.; Ornstein,~R. {Molecular dynamics study of time-correlated protein
  domain motions and molecular flexibility: Cytochrome P450BM-3}.
  \emph{Biophys. J.} \textbf{{1997}}, \emph{{73}}, {1147--1159}\relax
\mciteBstWouldAddEndPuncttrue
\mciteSetBstMidEndSepPunct{\mcitedefaultmidpunct}
{\mcitedefaultendpunct}{\mcitedefaultseppunct}\relax
\EndOfBibitem
\bibitem[Grant \latin{et~al.}({2006})Grant, Rodrigues, ElSawy, McCammon, and
  Caves]{bio3d}
Grant,~B.~J.; Rodrigues,~A. P.~C.; ElSawy,~K.~M.; McCammon,~J.~A.; Caves,~L.
  S.~D. {Bio3d: an R package for the comparative analysis of protein
  structures}. \emph{{Bioinformatics}} \textbf{{2006}}, \emph{{22}},
  {2695--2696}\relax
\mciteBstWouldAddEndPuncttrue
\mciteSetBstMidEndSepPunct{\mcitedefaultmidpunct}
{\mcitedefaultendpunct}{\mcitedefaultseppunct}\relax
\EndOfBibitem
\bibitem[Becker \latin{et~al.}(2018)Becker, Bhattiprolu, Gubens{\"a}k, and
  Zangger]{becker:2018}
Becker,~W.; Bhattiprolu,~K.~C.; Gubens{\"a}k,~N.; Zangger,~K. Investigating
  protein--ligand interactions by solution nuclear magnetic resonance
  spectroscopy. \emph{ChemPhysChem} \textbf{2018}, \emph{19}, 895\relax
\mciteBstWouldAddEndPuncttrue
\mciteSetBstMidEndSepPunct{\mcitedefaultmidpunct}
{\mcitedefaultendpunct}{\mcitedefaultseppunct}\relax
\EndOfBibitem
\end{mcitethebibliography}


\providecommand{\latin}[1]{#1}
\makeatletter
\providecommand{\doi}
  {\begingroup\let\do\@makeother\dospecials
  \catcode`\{=1 \catcode`\}=2 \doi@aux}
\providecommand{\doi@aux}[1]{\endgroup\texttt{#1}}
\makeatother
\providecommand*\mcitethebibliography{\thebibliography}
\csname @ifundefined\endcsname{endmcitethebibliography}
  {\let\endmcitethebibliography\endthebibliography}{}
\begin{mcitethebibliography}{0}
\providecommand*\natexlab[1]{#1}
\providecommand*\mciteSetBstSublistMode[1]{}
\providecommand*\mciteSetBstMaxWidthForm[2]{}
\providecommand*\mciteBstWouldAddEndPuncttrue
  {\def\EndOfBibitem{\unskip.}}
\providecommand*\mciteBstWouldAddEndPunctfalse
  {\let\EndOfBibitem\relax}
\providecommand*\mciteSetBstMidEndSepPunct[3]{}
\providecommand*\mciteSetBstSublistLabelBeginEnd[3]{}
\providecommand*\EndOfBibitem{}
\mciteSetBstSublistMode{f}
\mciteSetBstMaxWidthForm{subitem}{(\alph{mcitesubitemcount})}
\mciteSetBstSublistLabelBeginEnd
  {\mcitemaxwidthsubitemform\space}
  {\relax}
  {\relax}

\end{mcitethebibliography}

\end{document}


\begin{figure}[H]
\begin{center}
\includegraphics[width=0.9\textwidth]{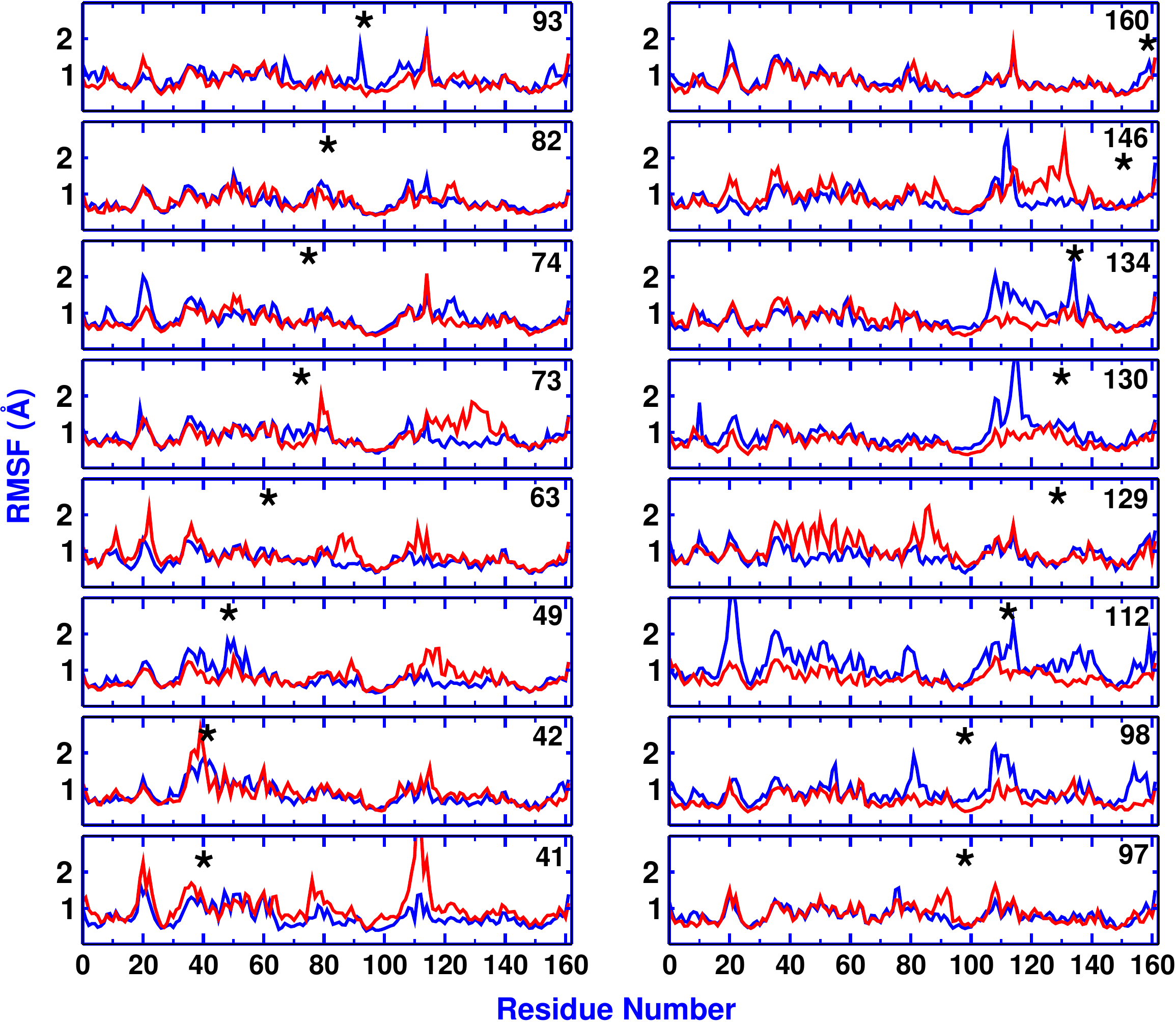}
\caption{Root mean squared fluctuations (RMSFs) for the C$_{\alpha}$
  atoms for ligand-free (blue) and PhCN-bound (red) Lysozyme with
  -N$_3$ attached to every alanine residues. The label in each panel
  refers to the alanine residue number which carries the azide label
  and the corresponding position of residue is indicated as asterisk
  above the RMSF trace.}
\label{sifig:rmsf}
\end{center}
\end{figure}

\begin{figure}[H]
\begin{center}
\includegraphics[width=0.9\textwidth,]{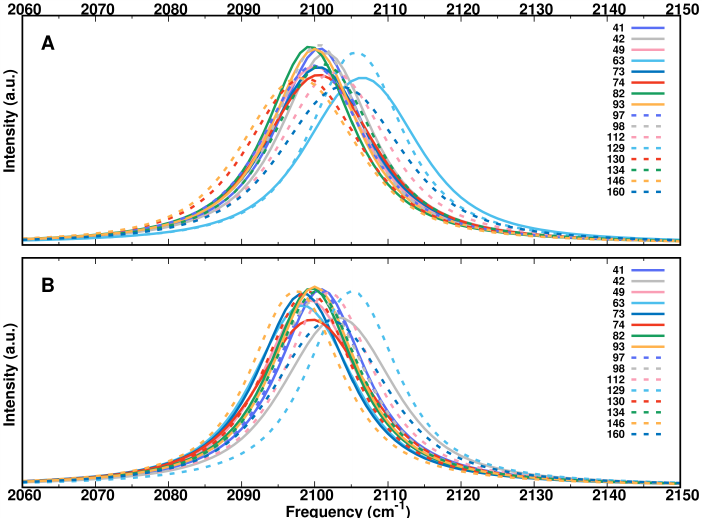}
\caption{1D lineshapes from INM for all 16 AlaN$_3$ residues for
  ligand-free (panel A) and ligand-bound (panel B) Lysozyme. The
  positions of all frequency maxima are compared in Figure 3.}
\label{sifig:spec}
\end{center}
\end{figure}

\begin{figure}[H]
\begin{center}
\includegraphics[width=0.9\textwidth,]{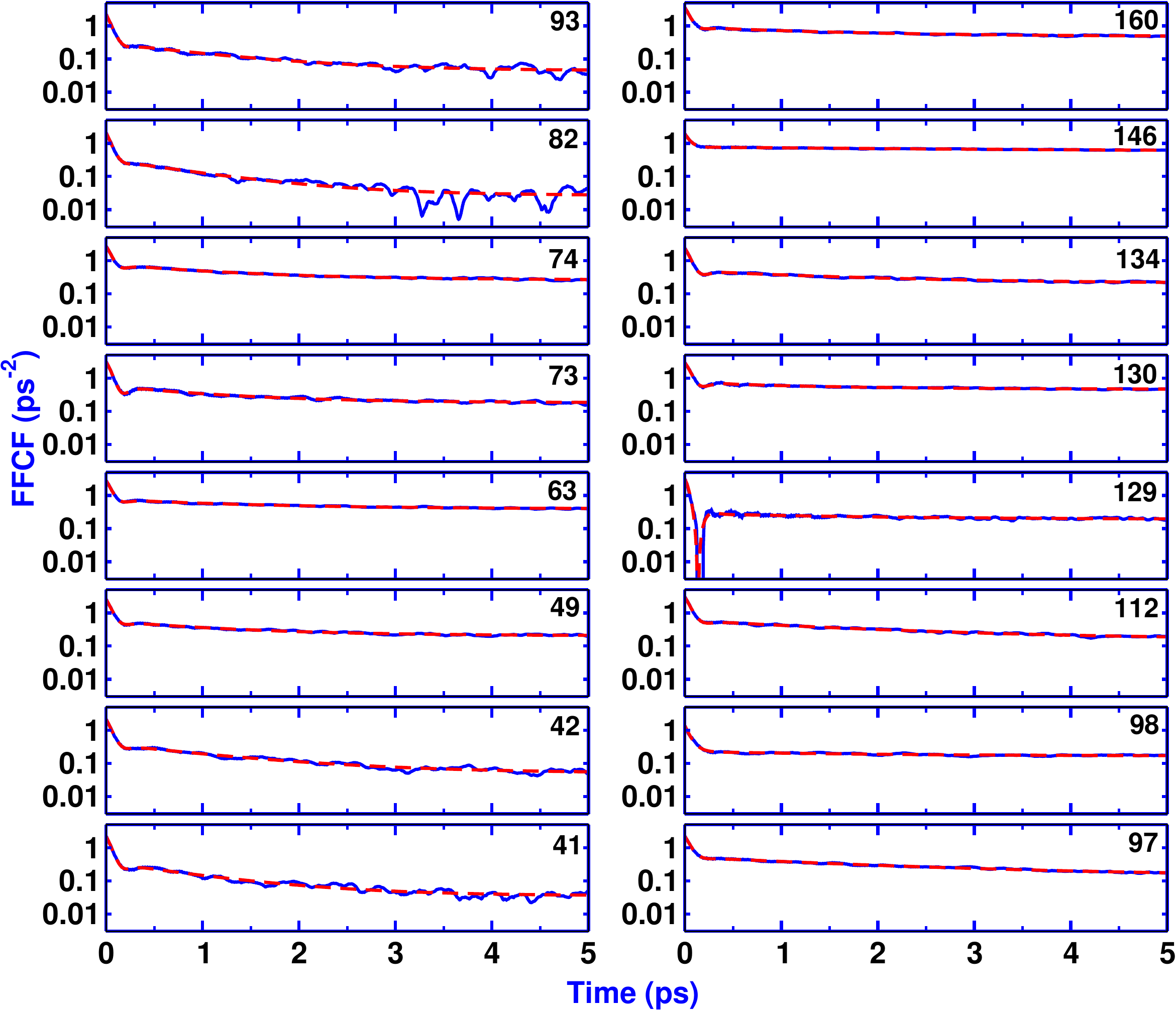}
\caption{FFCFs from correlating the instantaneous harmonic frequencies
  for all 16 AlaN$_3$ in Lysozyme. The labels in each panel refer to
  the alanine residue which carries the azide label. Black traces are
  the raw data and red dashed lines the fits to
  Eq. 1. The $y-$axis is logarithmic.}
\label{sifig:ffcffit_azo}
\end{center}
\end{figure}

\begin{figure}[H]
\begin{center}
\includegraphics[width=0.9\textwidth,]{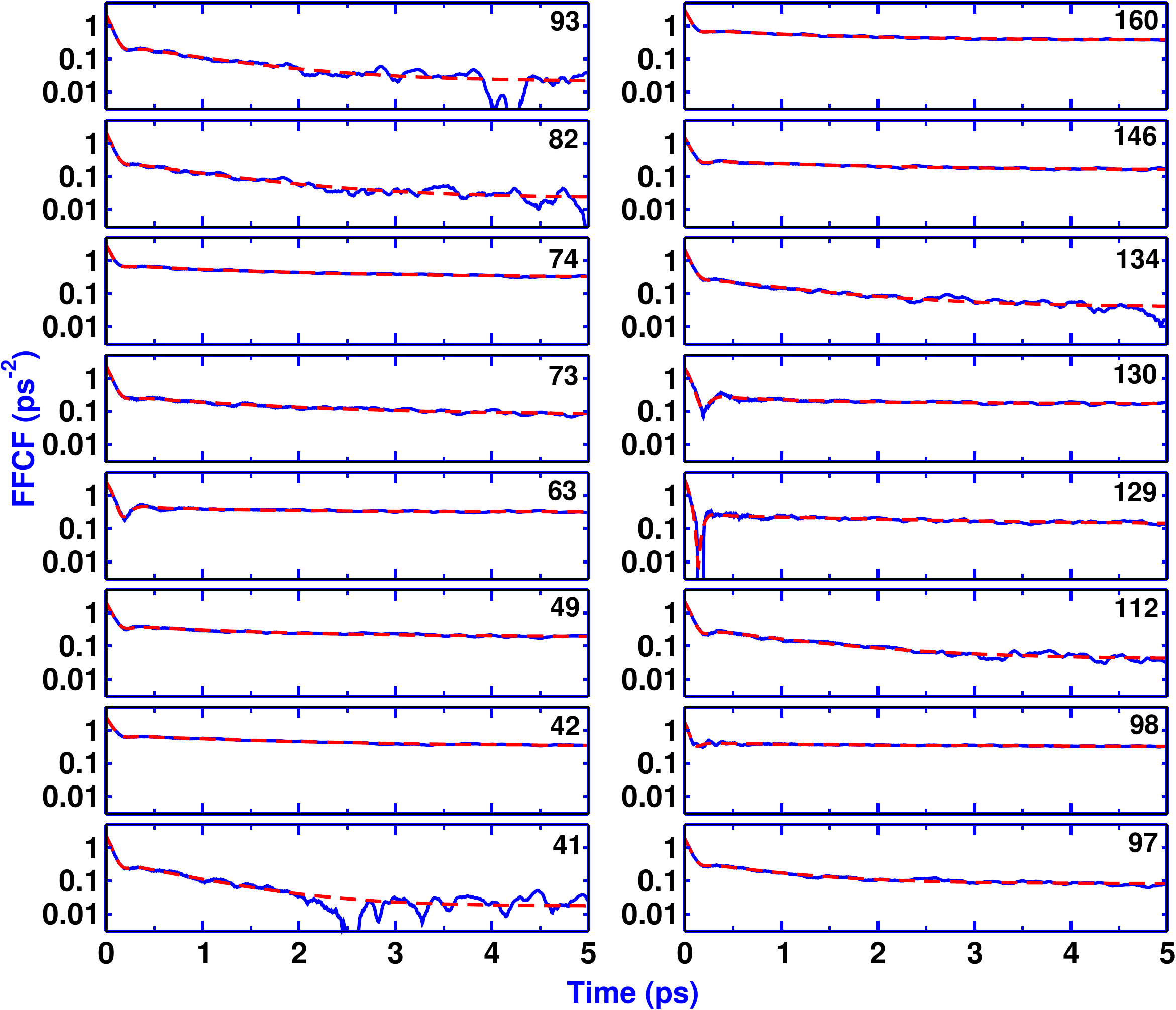}
\caption{FFCFs from correlating the instantaneous harmonic frequencies
  for all 16 AlaN$_3$ in Lysozyme with PhCN in the cavity. The labels in each panel refer to
  the alanine residue which carries the azide label. Black traces are
  the raw data and red dashed lines the fits to
  Eq. 1. The $y-$axis is logarithmic.}
\label{sifig:ffcffit_benzazo}
\end{center}
\end{figure}

\begin{table}[H]
\footnotesize
\centering
\caption{Parameters obtained from fitting the FFCF to
  Eq. 1 for INM frequencies for all different AlaN$_3$
  residues in lysozyme. Average frequency $\langle\omega\rangle$ of
  the asymmetric stretch in cm$^{-1}$, the amplitudes $a_1$ to $a_3$
  in ps$^{-2}$, the decay times $\tau_1$ to $\tau_3$ in ps, the
  parameter $\gamma$ in ps$^{-1}$ and the static term $\Delta_0^2$ in ps$^{-2}$.}  
 \centering
\begin{tabular}{|r|c|crc|cc|c|}
\hline\hline
\multicolumn{8}{c}{LysN$_3$} \\
\hline\hline
Res& $\langle\omega\rangle$ & $a_{1}$ & $\gamma$ &$\tau_{1}$ &$a_{2}$ &$\tau_{2}$ & $\Delta_0^2$\\
\hline
\textbf{41} &2099.69 & 1.86 &  11.23   & 0.068   & 0.32   & 0.93   &  0.03    \\
\hline
\textbf{42} & 2100.53   &  1.76  & 10.21   & 0.069   & 0.33   & 1.17   & 0.05   \\
\hline
\textbf{49} & 2099.02  &  2.00  & 10.67   & 0.069   & 0.36   & 1.18   & 0.20   \\
\hline
\textbf{63} & 2105.49   & 2.05 & 11.71   & 0.069   & 0.37   & 1.51  & 0.39   \\
\hline
\textbf{73} & 2099.56   & 2.49 & 13.04  & 0.078   & 0.40   & 1.05   & 0.18   \\
\hline
\textbf{74} & 2099.59   & 1.94 & 12.80   & 0.072   & 0.52   & 1.21   & 0.26   \\
\hline
\textbf{82} & 2098.36   & 1.76 & 9.15   & 0.062   & 0.30   & 0.90   & 0.02   \\
\hline
\textbf{93} & 2099.00   & 1.92 & 8.88  & 0.065   & 0.27 & 1.05   & 0.04   \\
\hline
\textbf{97} & 2098.86   & 1.70  & 7.96  & 0.067   & 0.39   & 2.28   &  0.13  \\
\hline
\textbf{98} & 2099.47   & 1.15 & 0.0   &0.057    & 0.06  & 2.04   & 0.16   \\
\hline
\textbf{112} & 2101.62   & 2.40 & 9.41   & 0.072   & 0.44   & 1.99   & 0.15   \\
\hline
\textbf{129} & 2104.69   & 2.83 & 18.08   & 0.066   & 0.09   & 1.98   & 0.18   \\
\hline
\textbf{130} & 2097.57   & 2.19 & 12.36   & 0.080   & 0.28   & 1.36 & 0.45   \\
\hline
\textbf{134} & 2100.17   & 1.87 & 11.55   & 0.074   & 0.29   & 1.79   & 0.20   \\
\hline
\textbf{146} & 2096.84   & 1.16 &6.15   &0.057    &0.24    & 5.13   & 0.52   \\
\hline
\textbf{160} & 2102.67  & 2.64 & 10.20  & 0.065   & 0.45   & 1.59   & 0.47   \\
\hline\hline
\multicolumn{8}{c}{LysN$_3$--PhCN} \\
\hline\hline
\textbf{41} &2099.90 & 1.77 &  11.78   & 0.068   & 0.38   & 0.70   &  0.01    \\
\hline
\textbf{42} & 2102.22   &  1.66  & 10.28   & 0.074   & 0.40   & 1.74   & 0.32   \\
\hline
\textbf{49} & 2098.31  &  1.56  & 10.52   & 0.072   & 0.23   & 1.34   & 0.19   \\
\hline
\textbf{63} & 2097.35   & 1.99 & 13.90   & 0.094   & 0.16   & 1.26   & 0.31   \\
\hline
\textbf{73} & 2097.38   & 1.94 & 8.92   & 0.067   & 0.22   & 1.40   & 0.07   \\
\hline
\textbf{74} & 2098.61   & 2.12 & 10.80   & 0.064   & 0.44   & 1.58   & 0.31   \\
\hline
\textbf{82} & 2098.62   & 1.76 & 9.10   & 0.063   & 0.28   & 0.96   & 0.02   \\
\hline
\textbf{93} & 2099.00   & 1.80 & 10.00  & 0.067   & 0.27 & 0.88   & 0.02   \\
\hline
\textbf{97} & 2099.65   & 1.46  & 10.88  & 0.064   & 0.31   & 0.81   &  0.08  \\
\hline
\textbf{98} & 2099.44   & 1.23 & 18.07   &0.054    & 0.09  & 1.96   & 0.32   \\
\hline
\textbf{112} & 2100.91   & 1.79 & 11.01   & 0.074   & 0.32   & 1.01   & 0.04   \\
\hline
\textbf{129} & 2104.11   & 2.55 & 17.75   & 0.066   & 0.14   & 2.78   & 0.12   \\
\hline
\textbf{130} & 2098.13   & 1.64 & 13.09   & 0.090   & 0.13   & 1.00 &0.17   \\
\hline
\textbf{134} & 2099.16   & 1.76 & 8.28   & 0.063   & 0.29   & 1.03   & 0.03   \\
\hline
\textbf{146} & 2096.41   & 1.17 &10.61   &0.068    &0.15    & 1.61   & 0.15   \\
\hline
\textbf{160} & 2101.14  & 2.08 & 11.65  & 0.069   & 0.42   & 1.29   & 0.37   \\
\hline\hline
\end{tabular}
\label{sitab:ffcffit}
\end{table}

\begin{figure}[H]
\begin{center}
\includegraphics[width=0.5\textwidth]{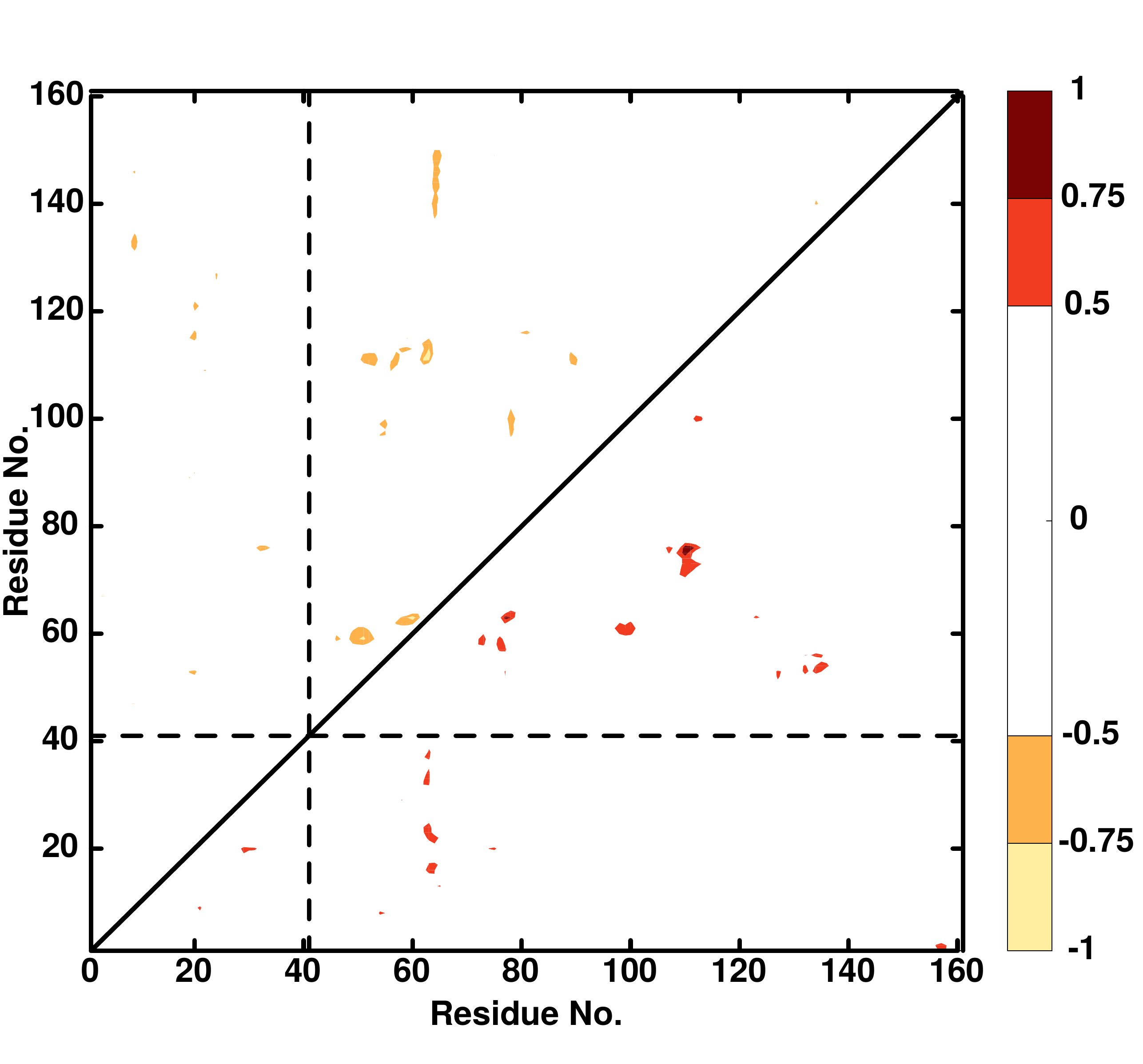}
\caption{Difference dynamic cross correlation maps ($\Delta$DCCM)
  between Ala41N$_3$ and Ala41N$_3$-PhCN. Positive correlations are in the lower right triangle, negative correlations in the upper left triangle. Only correlation coefficients with an absolute value greater than 0.5 are displayed.}
\label{sifig:41}
\end{center}
\end{figure}

\begin{figure}[H]
\begin{center}
\includegraphics[width=0.5\textwidth]{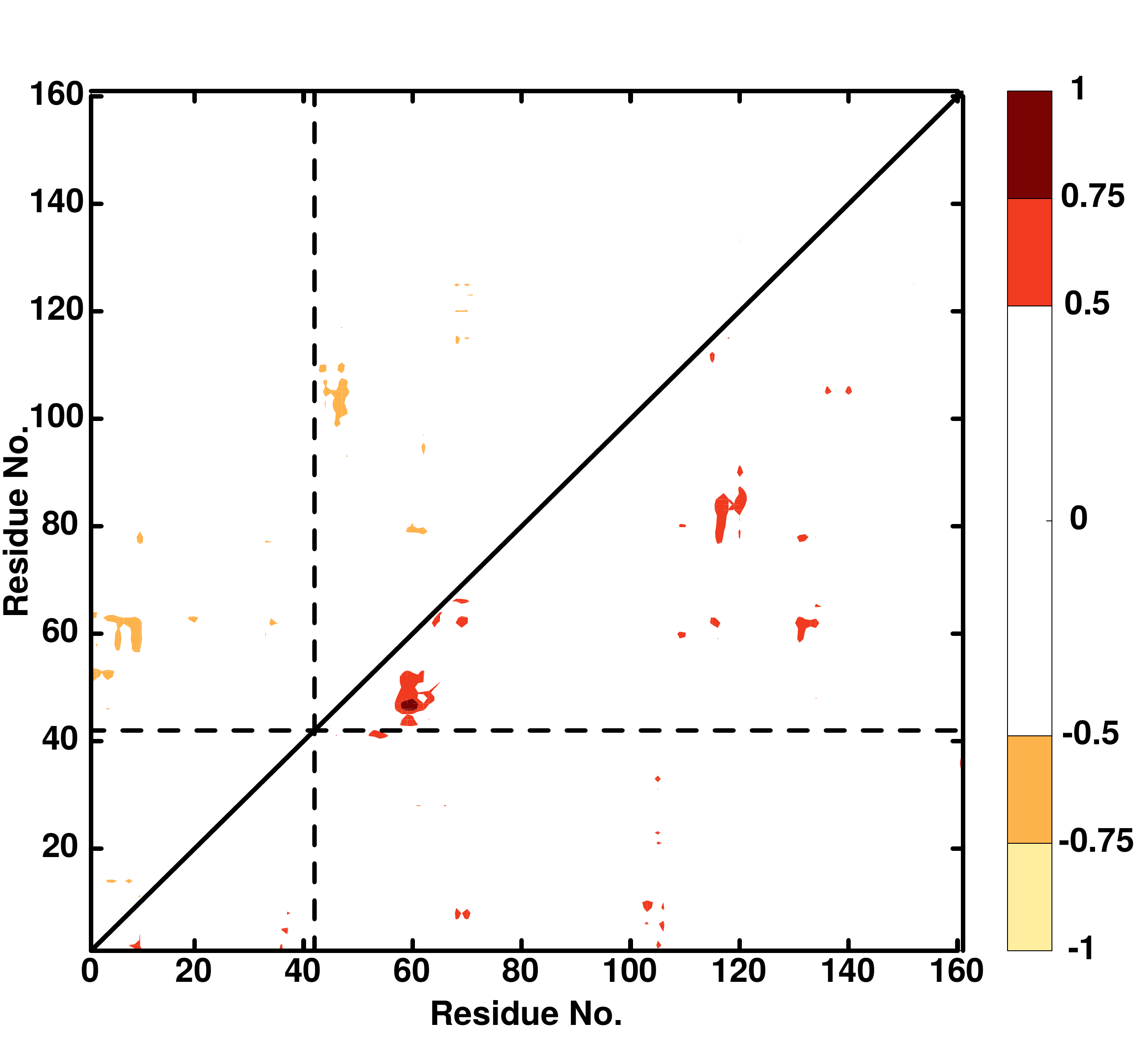}
\caption{Difference dynamic cross correlation maps ($\Delta$DCCM)
  between  Ala42N$_3$ and Ala42N$_3$-PhCN. Positive correlations are in the lower right triangle, negative correlations in the upper left triangle. Only correlation coefficients with an absolute value greater than 0.5 are displayed.}
\label{sifig:42}
\end{center}
\end{figure}

\begin{figure}[H]
\begin{center}
\includegraphics[width=0.5\textwidth]{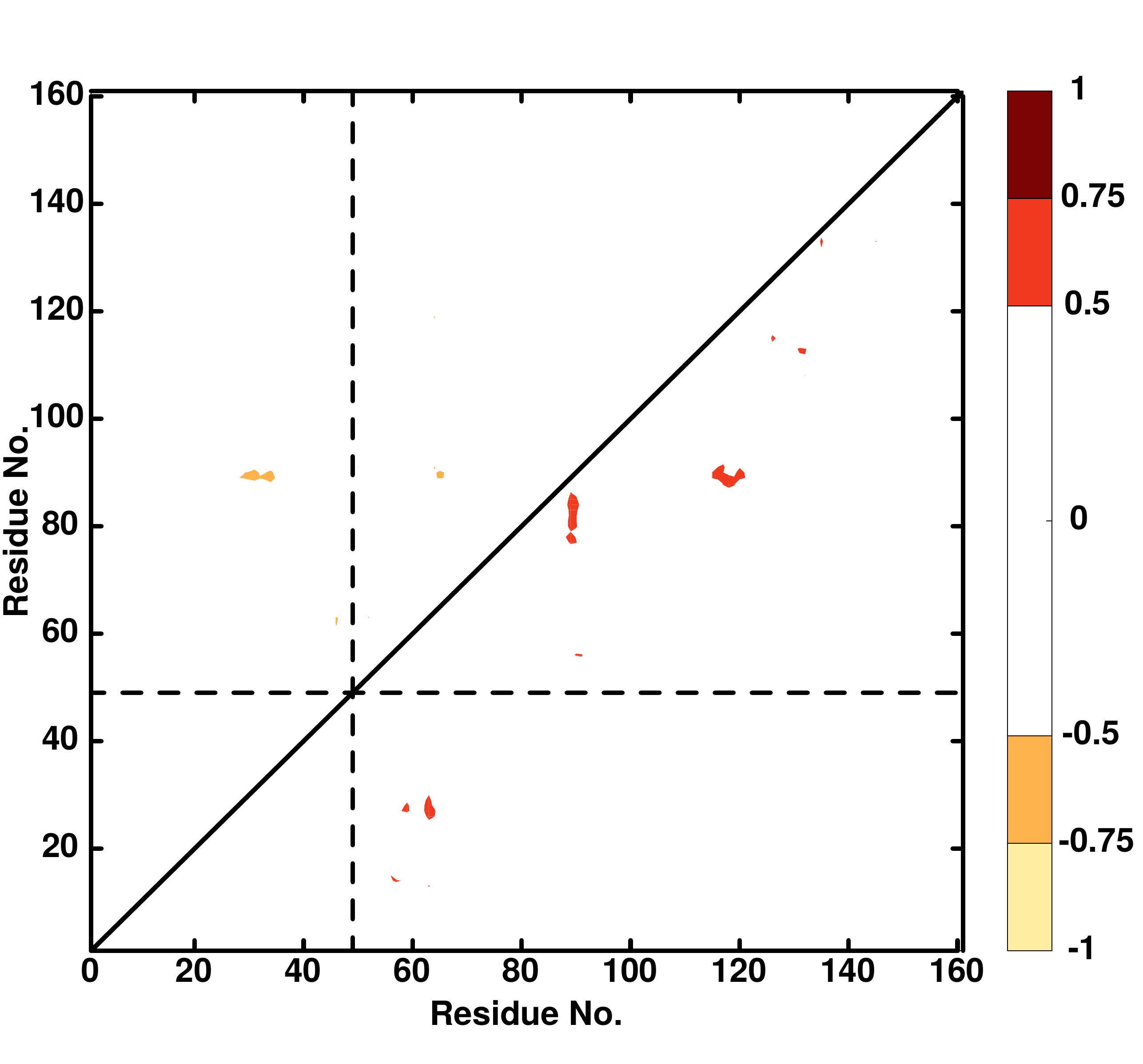}
\caption{Difference dynamic cross correlation maps ($\Delta$DCCM)
  between Ala49N$_3$ and Ala49N$_3$-PhCN. Positive correlations are in the lower right triangle, negative correlations in the upper left triangle. Only correlation coefficients with an absolute value greater than 0.5 are displayed.}
\label{sifig:49}
\end{center}
\end{figure}

\begin{figure}[H]
\begin{center}
\includegraphics[width=0.5\textwidth]{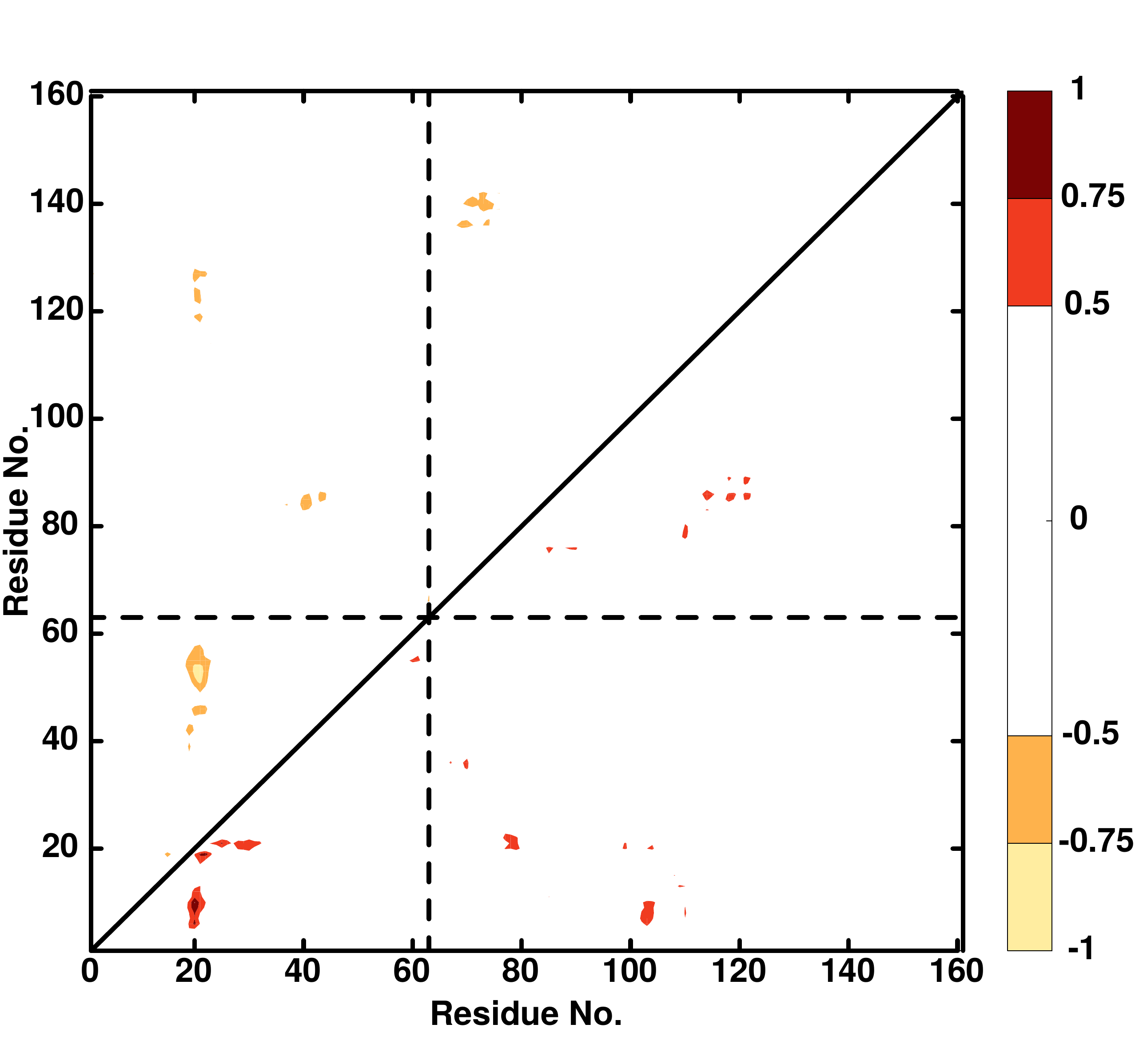}
\caption{Difference dynamic cross correlation maps ($\Delta$DCCM)
  between Ala63N$_3$ and Ala63N$_3$-PhCN. Positive correlations are in the lower right triangle, negative correlations in the upper left triangle. Only correlation coefficients with an absolute value greater than 0.5 are displayed.}
\label{sifig:63}
\end{center}
\end{figure}

\begin{figure}[H]
\begin{center}
\includegraphics[width=0.5\textwidth]{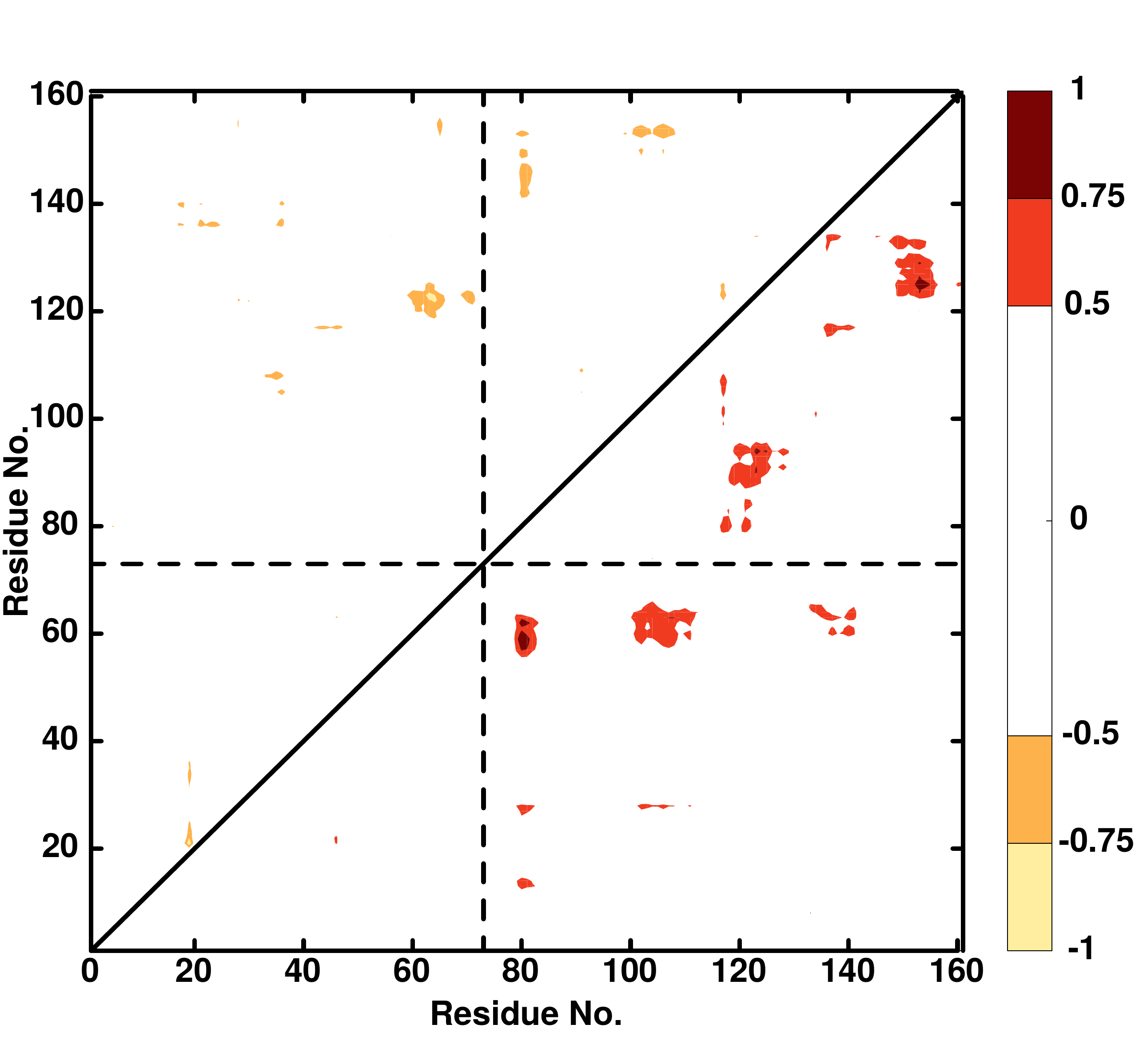}
\caption{Difference dynamic cross correlation maps ($\Delta$DCCM)
  between Ala73N$_3$ and Ala73N$_3$-PhCN. Positive correlations are in the lower right triangle, negative correlations in the upper left triangle. Only correlation coefficients with an absolute value greater than 0.5 are displayed.}
\label{sifig:73}
\end{center}
\end{figure}

\begin{figure}[H]
\begin{center}
\includegraphics[width=0.5\textwidth]{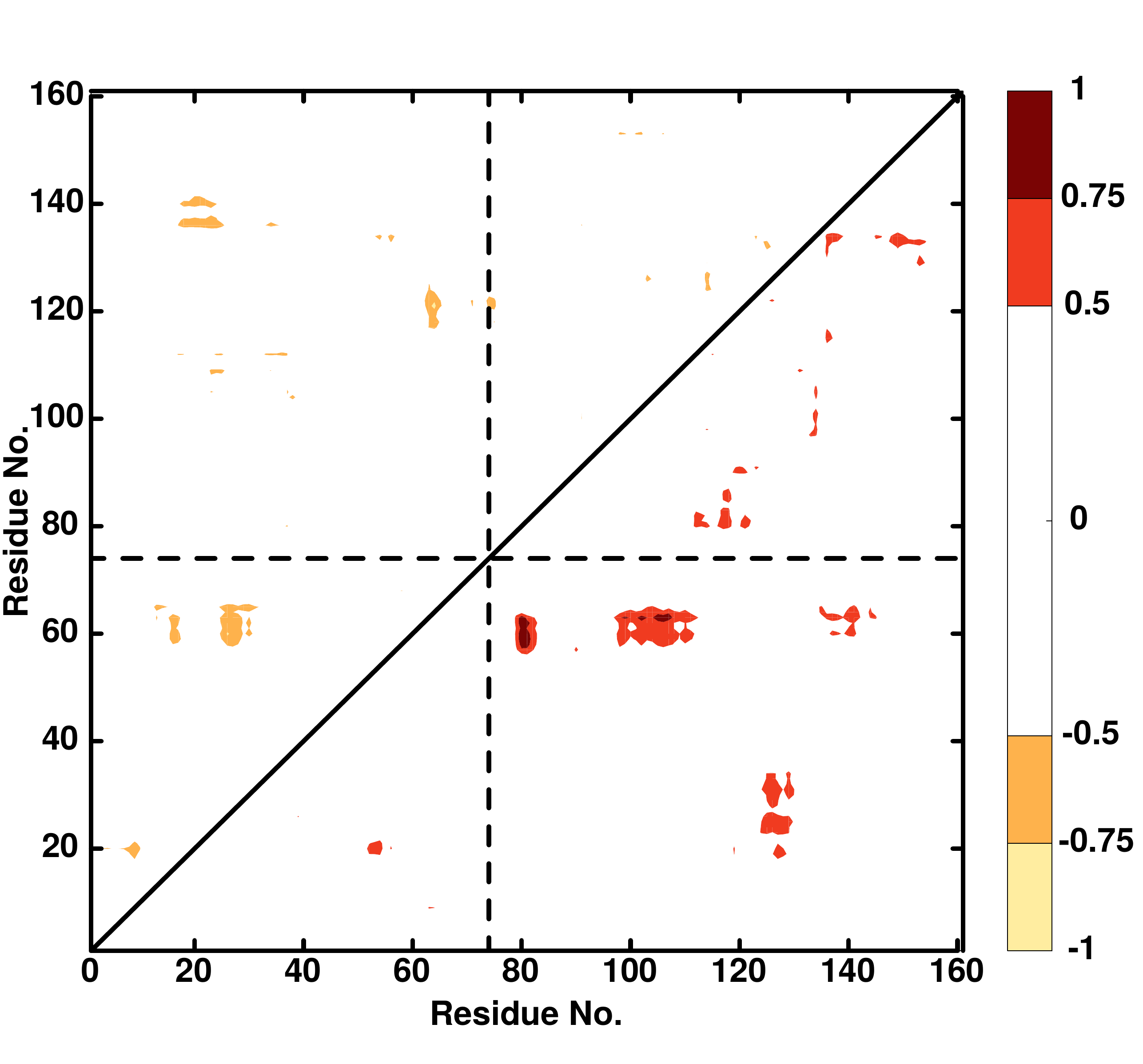}
\caption{Difference dynamic cross correlation maps ($\Delta$DCCM)
  between Ala74N$_3$ and Ala74N$_3$-PhCN. Positive correlations are in the lower right triangle, negative correlations in the upper left triangle. Only correlation coefficients with an absolute value greater than 0.5 are displayed.}
\label{sifig:74}
\end{center}
\end{figure}

\begin{figure}[H]
\begin{center}
\includegraphics[width=0.5\textwidth]{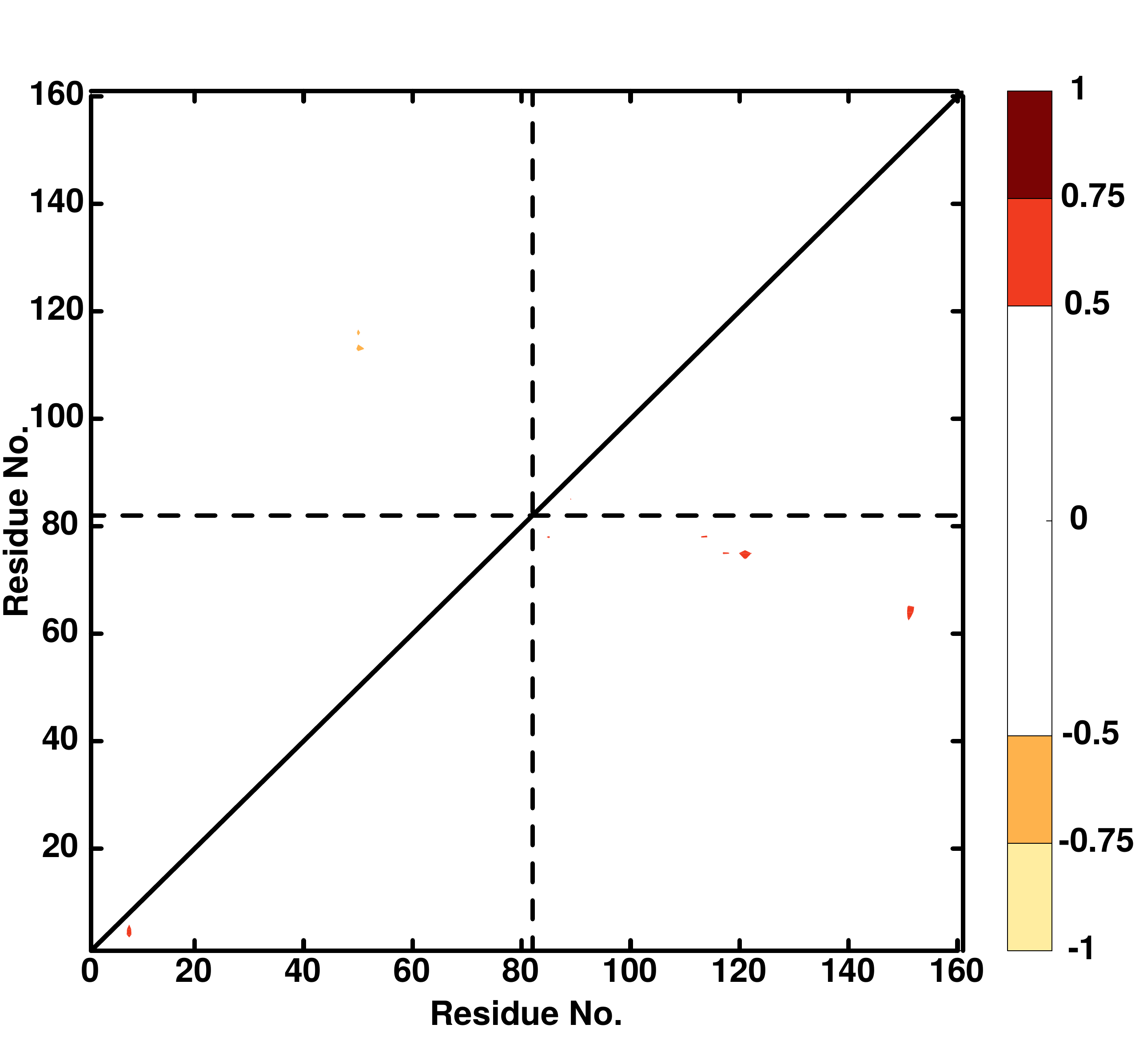}
\caption{Difference dynamic cross correlation maps ($\Delta$DCCM)
  between Ala82N$_3$ and Ala82N$_3$-PhCN. Positive correlations are in the lower right triangle, negative correlations in the upper left triangle. Only correlation coefficients with an absolute value greater than 0.5 are displayed.}
\label{sifig:82}
\end{center}
\end{figure}

\begin{figure}[H]
\begin{center}
\includegraphics[width=0.5\textwidth]{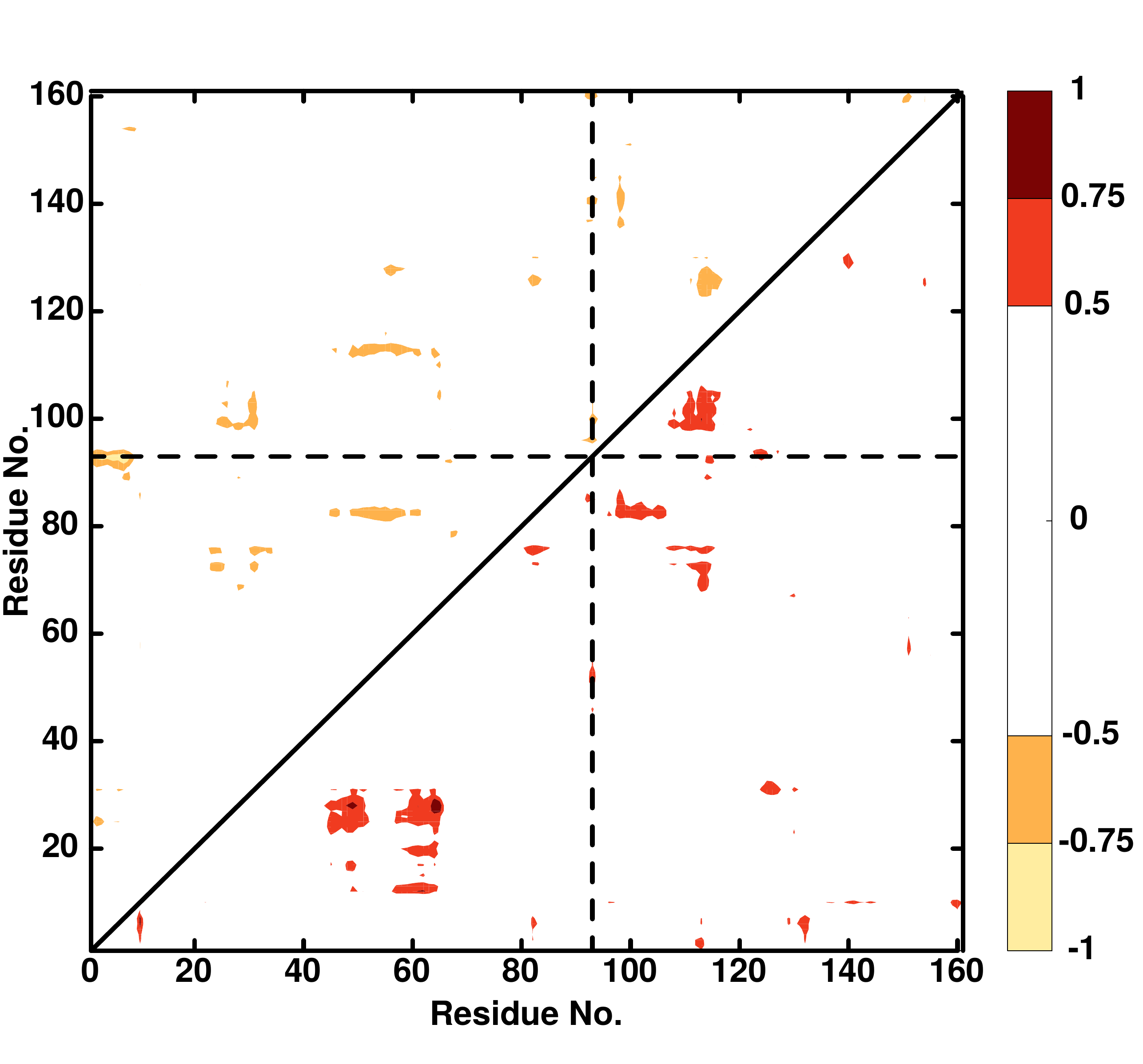}
\caption{Difference dynamic cross correlation maps ($\Delta$DCCM)
  between Ala93N$_3$ and Ala93N$_3$-PhCN. Positive correlations are in the lower right triangle, negative correlations in the upper left triangle. Only correlation coefficients with an absolute value greater than 0.5 are displayed.}
\label{sifig:93}
\end{center}
\end{figure}

\begin{figure}[H]
\begin{center}
\includegraphics[width=0.5\textwidth]{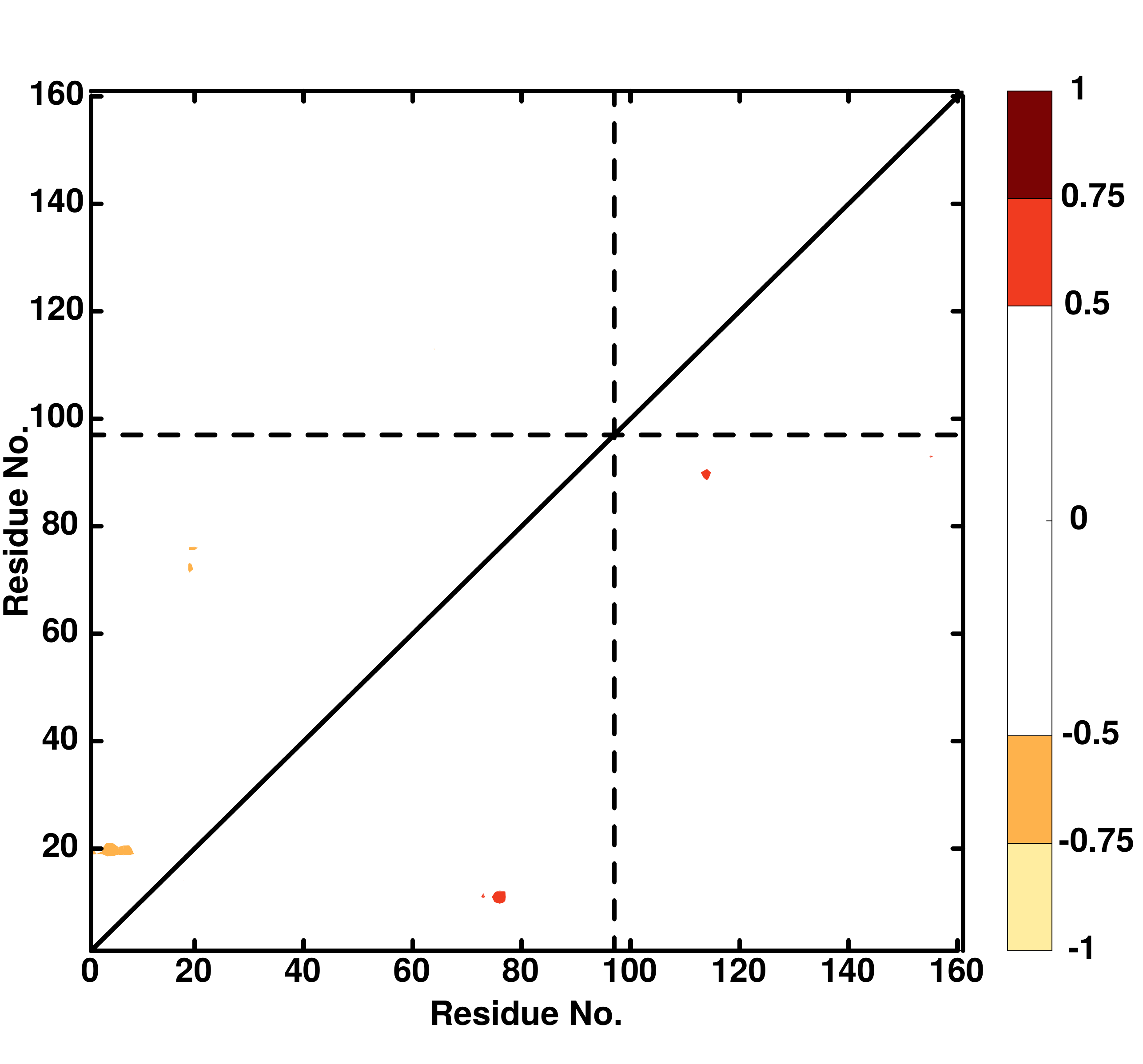}
\caption{Difference dynamic cross correlation maps ($\Delta$DCCM)
  between Ala97N$_3$ and Ala97N$_3$-PhCN. Positive correlations are in the lower right triangle, negative correlations in the upper left triangle. Only correlation coefficients with an absolute value greater than 0.5 are displayed.}
\label{sifig:97}
\end{center}
\end{figure}

\begin{figure}[H]
\begin{center}
\includegraphics[width=0.5\textwidth]{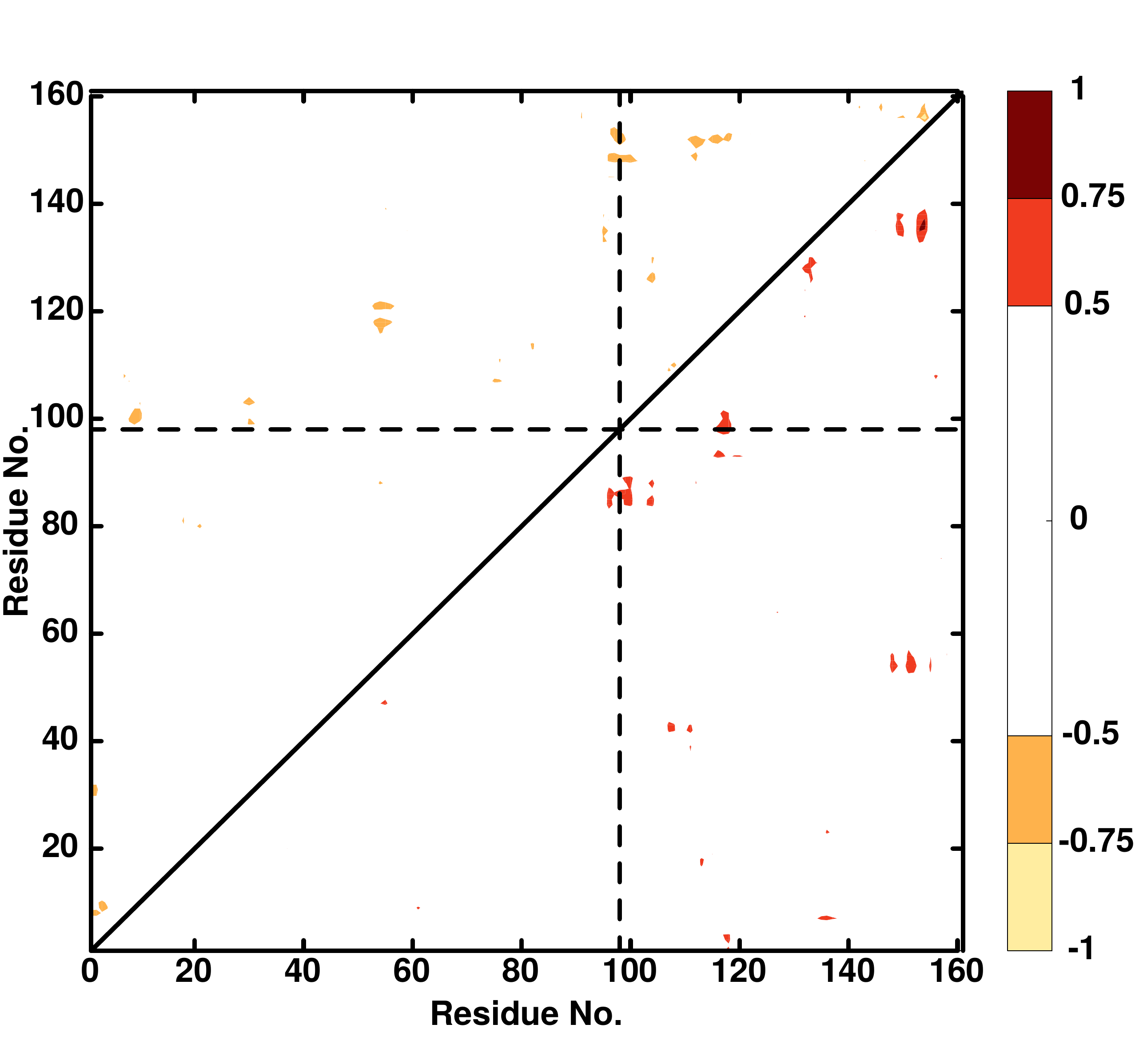}
\caption{Difference dynamic cross correlation maps ($\Delta$DCCM)
  between Ala98N$_3$ and Ala98N$_3$-PhCN. Positive correlations are in the lower right triangle, negative correlations in the upper left triangle. Only correlation coefficients with an absolute value greater than 0.5 are displayed.}
\label{sifig:98}
\end{center}
\end{figure}

\begin{figure}[H]
\begin{center}
\includegraphics[width=0.5\textwidth]{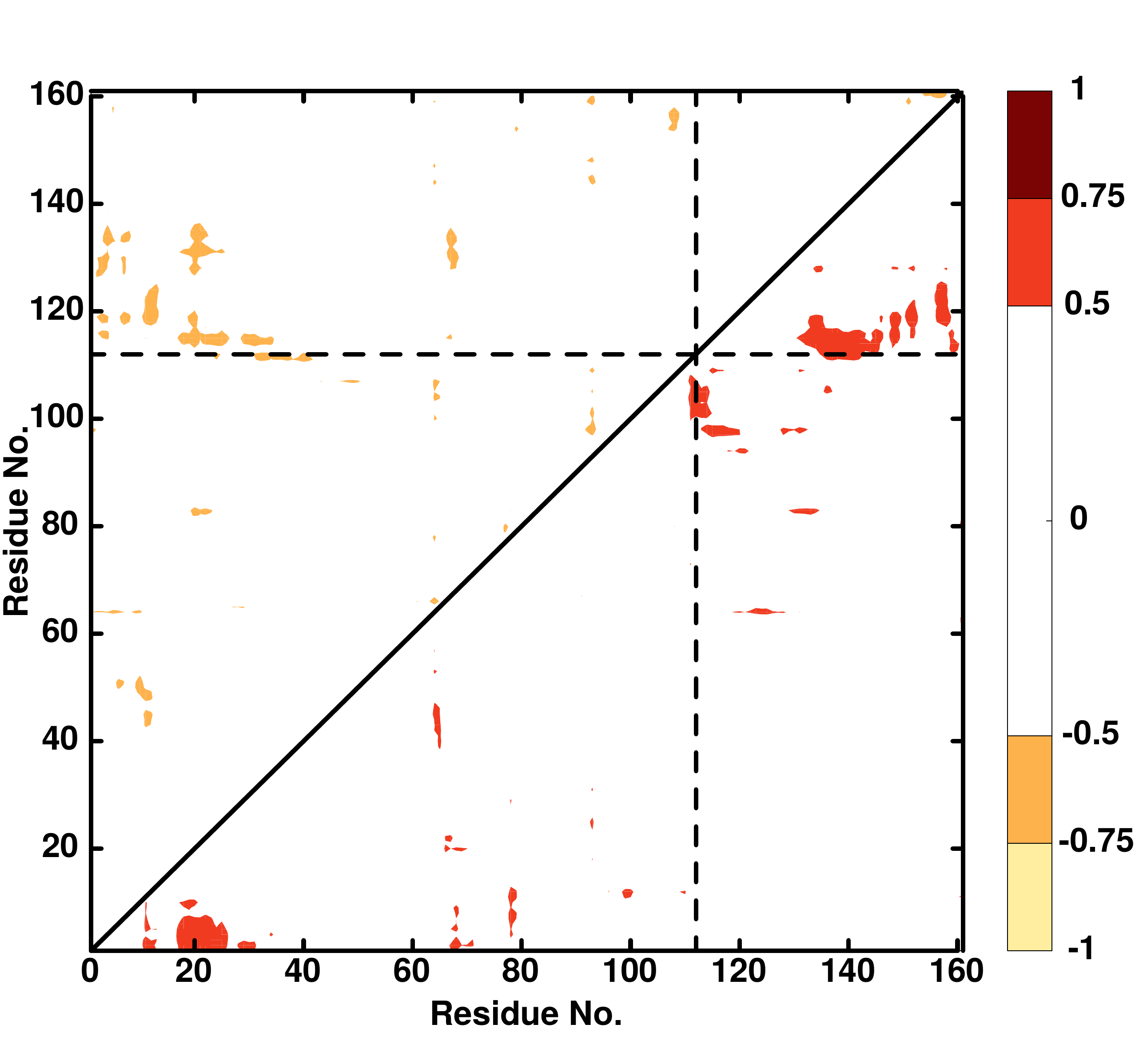}
\caption{Difference dynamic cross correlation maps ($\Delta$DCCM)
  between Ala112N$_3$ and Ala112N$_3$-PhCN. Positive correlations are in the lower right triangle, negative correlations in the upper left triangle. Only correlation coefficients with an absolute value greater than 0.5 are displayed.}
\label{sifig:112}
\end{center}
\end{figure}

\begin{figure}[H]
\begin{center}
\includegraphics[width=0.5\textwidth]{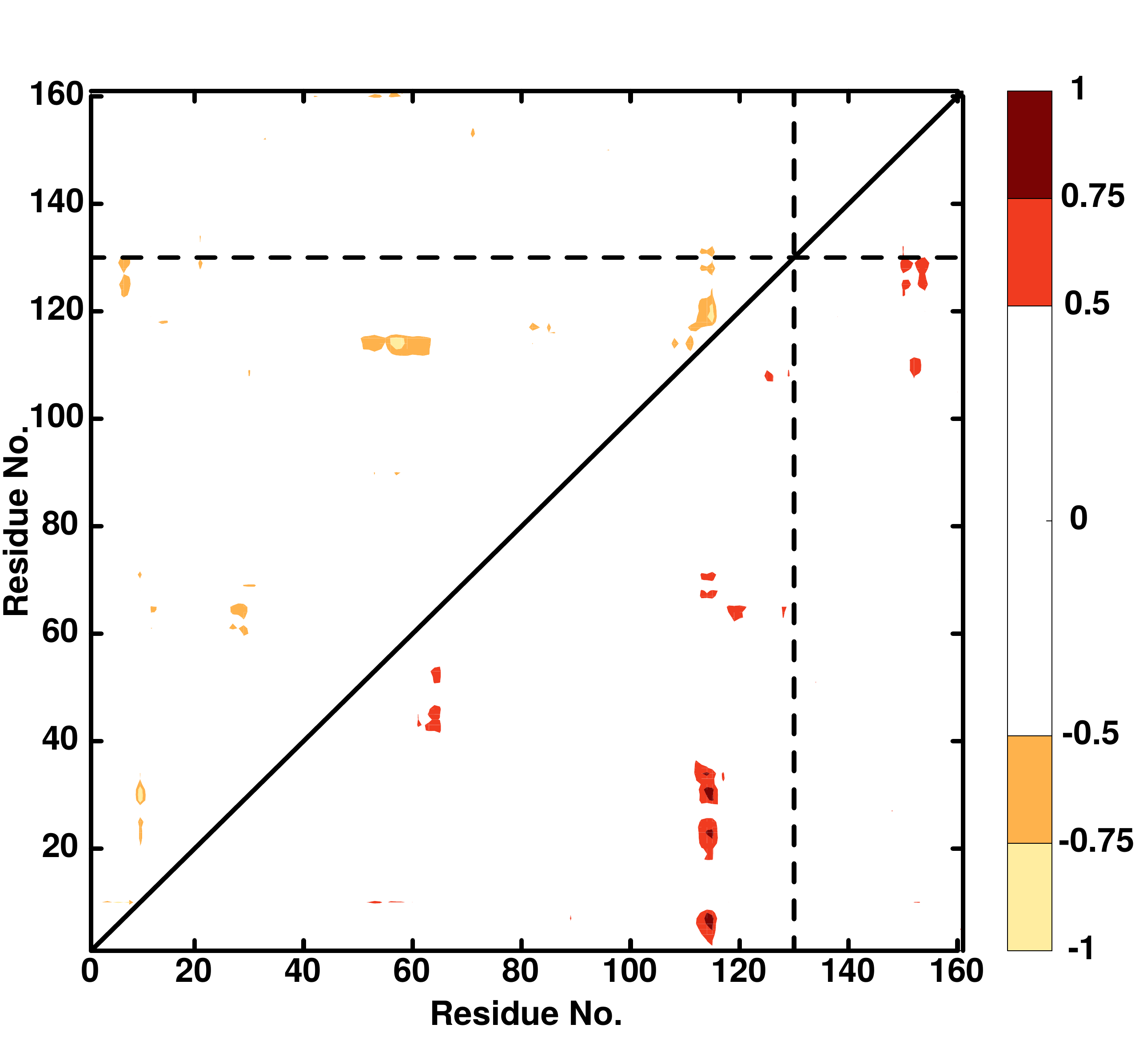}
\caption{Difference dynamic cross correlation maps ($\Delta$DCCM)
  between Ala130N$_3$ and Ala130N$_3$-PhCN. Positive correlations are in the lower right triangle, negative correlations in the upper left triangle. Only correlation coefficients with an absolute value greater than 0.5 are displayed.}
\label{sifig:130}
\end{center}
\end{figure}

\begin{figure}[H]
\begin{center}
\includegraphics[width=0.5\textwidth]{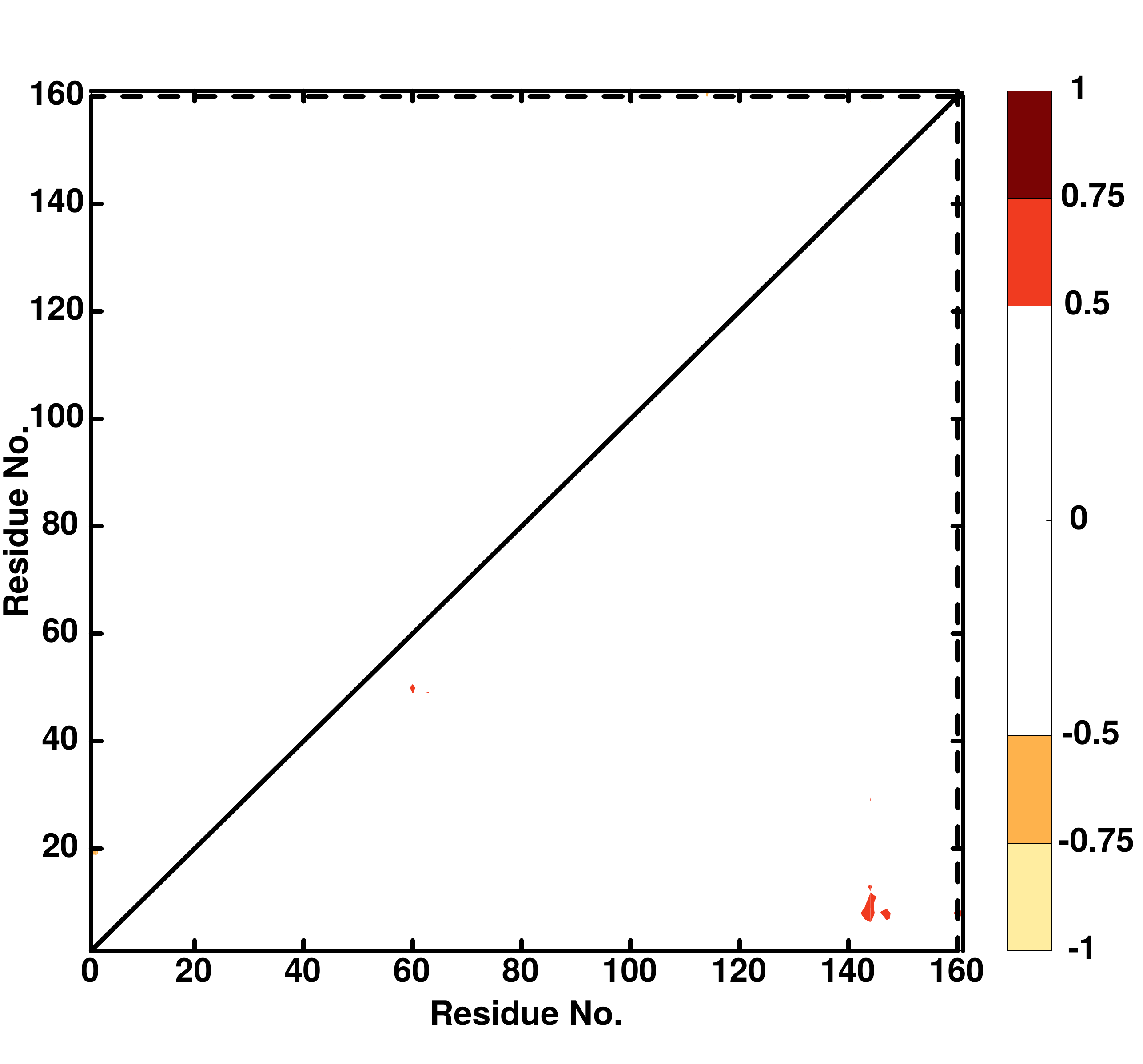}
\caption{Difference dynamic cross correlation maps ($\Delta$DCCM)
  between  Ala160N$_3$ and Ala160N$_3$-PhCN. Positive correlations are in the lower right triangle, negative correlations in the upper left triangle. Only correlation coefficients with an absolute value greater than 0.5 are displayed.}
\label{sifig:160}
\end{center}
\end{figure}